\documentclass[aps,prl,twocolumn,amsmath,amssymb,showpacs]{revtex4-1}
\usepackage{amsmath}
\usepackage{amssymb}
\usepackage{graphicx}
\usepackage{float}
\usepackage{subfigure}
\usepackage{color}
\usepackage{hyperref}
\usepackage{comment}
\usepackage{makecell}

\begin{document}
\title{Kardar-Parisi-Zhang universality class in ($d+1$)-dimensions}
\author{Tiago J. Oliveira}
\email{tiago@ufv.br}
\affiliation{Departamento de F\'isica, Universidade Federal de Vi\c cosa, 36570-900, Vi\c cosa, MG, Brazil}
\date{\today}

\begin{abstract}
The determination of the exact exponents of the KPZ class in any substrate dimension $d$ is one of the most important open issues in Statistical Physics. Based on the behavior of the dimensional variation of some exact exponent differences for other growth equations, I find here that the KPZ growth exponents (related to the temporal scaling of the fluctuations) are given by $\beta_d = \frac{7}{8d+13}$. These exponents present an excellent agreement with the most accurate estimates for them in the literature. Moreover, they are confirmed here through extensive Monte Carlo simulations of discrete growth models and real-space renormalization group (RG) calculations for directed polymers in random media (DPRM), up to $d=15$. The left-tail exponents of the probability density functions for the DPRM energy provide another striking verification of the analytical result above.
\end{abstract}

\maketitle

%\section{Introduction}
%\label{secIntro}

The Kardar-Parisi-Zhang (KPZ) \cite{KPZ} equation
\begin{equation}
 \frac{\partial h(\vec{x},t)}{\partial t} = \nu \nabla^2 h + \frac{\lambda}{2} (\nabla h)^2 + \eta(\vec{x},t),
\label{eqKPZ}
\end{equation}
where $\eta(\vec{x},t)$ is a white noise with $\langle\eta(\vec{x},t)\rangle = 0$ and $\langle\eta(\vec{x},t)\eta(\vec{x}',t') \rangle = 2 B \delta^d(\vec{x}-\vec{x}')\delta(t-t')$, describes the nonequilibrium dynamics of the field $h(\vec{x},t)$ and is related to so many systems \cite{barabasi,healy95,KrugAdv,Kriecherbauer2010} that it is considered as the nonequilibrium counterpart of the Ising class for equilibrium. 
Coincidentally, it took 24 years, since the introduction of the Ising model (by Lenz!), for the publication of the Onsager's solution in dimension $D=2$ \cite{Brush67} and the same time window until the first exact solutions of the KPZ equation in $d=1$ (or $D=d+1=2$), yielding the probability density functions (PDFs), $P(h)$, for the transient regime \cite{Sasamoto2010,Amir}.

These similarities end when one goes to higher dimensions, since the Ising exponents assume mean-field values for $D \ge 4$ (with relevant logarithmic corrections at the upper critical dimension $D_u = 4$) \cite{Cardy}, whereas for the KPZ class no exponent is exactly known for $d>1$ and the existence of a finite upper critical dimension $d_u$ is still a topic of debate. For instance, most of the studies based on mode-coupling theory \cite{Moore,Cates,Bray,Claudin} and field theoretical approaches \cite{Healy90,Kinzelbach,Fogedby2} have predicted $2.8 \leqslant d_u \leqslant 4$. Some real space renormalization group (RG) approaches indicate that $d_u$ is finite and larger than four \cite{Canet,Kloss}, while others suggest that $d_u = \infty$ \cite{Perlsman0,Castellano1,Castellano2}. Numerical simulations of several KPZ models have also provided strong evidence that no finite/small $d_u$ exists for the KPZ class \cite{Ala1,Ala2,Marinari02,Perlsman2,Perlsman1,Geza10,Parisi13,Fernando,JMKim13,Alves14,Kim2019,HHTake2015,Alves16}. 
The related scenario for the KPZ exponents is not too much better. In $d=2$, for example, the exponents estimated from different theoretical approaches \cite{Lassig,Moore,Fogedby} present significant discrepancies with the most accurate numerical estimates for them \cite{Kelling,Pagnani,Ismael22,Kelling18}; and a similar situation is observed in higher dimensions.

One of the main applications of the KPZ equation is in describing the evolution of the height field $h(\vec{x},t)$ of growing interfaces \cite{KPZ,barabasi,KrugAdv}. In general, when the growth is performed on an initially flat ($d$-dimensional) substrate of lateral size $L$, the squared interface width (or roughness) can be defined as $W_2(L,t) = \langle \bar{h^2} -  \bar{h}^2\rangle$, where $\bar{\cdot}$ and $\langle \cdot \rangle$ denote averages over the $L^d$ substrate sites and over $M$ different samples, respectively. During the transient growth regime, which exists while $t \ll t_c \sim L^{z_d}$, one has $W_2 \sim t^{2\beta_d}$ asymptotically. For $t \gg t_c$, $W_2$ stops increasing, attaining a steady state regime where its saturation value behaves asymptotically as $W_{2,s} \sim L^{2\alpha_d}$. Hence, the relevant exponents characterizing the system are $\alpha_d$ (the roughness), $\beta_d$ (the growth) and $z_d$ (the dynamic exponent), which are not independent at all, since the crossover scaling yields
\begin{equation}
 z_d = \frac{\alpha_d}{\beta_d}.
 \label{eqBeta}
\end{equation}
Thanks to a fluctuation-dissipation theorem, valid only for $d=1$, the KPZ exponents are exactly known in this dimension, being 
$\alpha_1=\frac{1}{2}$, $\beta_1=\frac{1}{3}$ and $z_1=\frac{3}{2}$ \cite{KPZ}. Moreover, from a tilting symmetry (or Galilean invariance) of the KPZ equation \cite{barabasi},
\begin{equation}
\alpha_d+z_d=2
\label{eqGI}
\end{equation}
is expected to hold for all $d$. In addition to Eqs. \ref{eqBeta} and \ref{eqGI}, a third (independent) scaling relation is required to fix the exponent values in general, but it is still unknown despite 36 years of efforts to find it. The aim of this Letter is to determine this missing relation and, then, the KPZ exponents for any $d > 1$.

In order to do this, I will start identifying some key properties of exponent differences for some exactly solvable growth equations. For example, the linear equations:
\begin{equation}
 \frac{\partial h(\vec{x},t)}{\partial t} = -(-1)^{\kappa} \nu_{\kappa} \nabla^{2\kappa} h + \eta(\vec{x},t),
\label{eqLinear}
\end{equation}
have the exponents $\alpha_d = (2\kappa-d)/2$ and $\beta_d = (2\kappa-d)/4\kappa$, where $\kappa =1,2,\ldots$ \cite{KrugAdv}. For $\kappa=1$, this gives the Edwards-Wilkinson (EW) \cite{EW} equation, which is the linear counterpart of the KPZ equation; while the $\kappa=2$ case is important, e.g., in the context of thin film deposition by molecular beam epitaxy (MBE) \cite{barabasi,KrugAdv}. Note that the upper critical dimensions (at which $\alpha_d=\beta_d=0$) of these linear classes are $d_u = 2\kappa$. Let us thus concentrate on some differences between their scaling exponents. First, note that the difference $\Gamma_d =\alpha_d - \beta_d$ follows the relation
\begin{equation}
\Gamma_{d+1} - \Gamma_d = - \beta_1.
\label{eqAlphaBeta1}
\end{equation}
Namely, the dimensional variation $G_{\Gamma} \equiv \Gamma_{d+1} - \Gamma_d$ does not depend on the dimensionality $d$. As an aside, I notice that the exponents $\alpha_d$ and $\beta_d$ change if one replaces the white noise $\eta$ in Eq. \ref{eqLinear} by a conserved noise or a `colored' (spatially- and/or temporally-correlated) noise \cite{barabasi,KrugAdv}, but in all cases Eq. \ref{eqAlphaBeta1} remains valid with $\beta_1 = (2\kappa-1)/4\kappa$.%{\color{red}[with $\langle\eta_k(\vec{x},t)\eta_k(\vec{x}',t') \rangle = -B\nabla^2 \delta^d(\vec{x}-\vec{x}')\delta(t-t')$, yielding $\alpha_d = (2\kappa -d-2)/2$ and $\beta_d = (2\kappa - d-2)/4\kappa$] or a `colored' (spatially- and/or temporally-correlated) noise [$\langle\eta_c(\vec{x},t)\eta_c(\vec{x}',t') \rangle \sim |\vec{x}-\vec{x}'|^{2\psi-d} |t-t'|^{2\phi-1}$, where $\alpha_d = (2\kappa -d)/2 + \psi + 2\kappa \phi$ and $\beta_d = (2\kappa -d)/4\kappa + \psi/2\kappa + \phi$] \cite{barabasi,KrugAdv}. However, in all cases Eq. \ref{eqAlphaBeta1} remains valid with $\beta_1 = (2\kappa-1)/4\kappa$, as in the white-noise case.}

It turns out that the apparent universality of the behavior in Eq. \ref{eqAlphaBeta1} breaks down for nonlinear interfaces. For instance, the nonlinear theory for MBE growth, i.e., the Villain-Lai-Das Sarma (VLDS) \cite{Villain,LDS} equation:% --- which is obtained by adding the term $\lambda_2 \nabla^2(\nabla h)^2$ in the right-hand-side of Eq. \ref{eqLinear} for $\kappa=2$ ---
\begin{equation}
 \frac{\partial h(\vec{x},t)}{\partial t} = -\nu_2 \nabla^{4} h + \lambda_1 \nabla^2(\nabla h)^2 + \eta(\vec{x},t),
\label{eqVLDS}
\end{equation}
has the (one loop) exponents $\alpha_d = (4-d)/3$ and $\beta_d = (4-d)/(8+d)$ \cite{barabasi}, which yield a variation $G_{\Gamma}$ dependent on $d$. Interestingly, however, $\Gamma_2 - \Gamma_1 = -\beta_2$ in this class. Hence, at least for the lowest (and physically relevant) dimensions, $G_{\Gamma}$ is given by a growth exponent, similarly to the linear systems. Notwithstanding, in the KPZ case things are more complicated, since Eq. \ref{eqAlphaBeta1} gives $\Gamma_2 = \Gamma_1 - \beta_1 = -1/6$, while one should have $\Gamma_2>0$, since $\alpha_2 > \beta_2$. Moreover, from the VLDS relation $\Gamma_2 = \Gamma_1 - \beta_2$, one gets $\alpha_2 = \Gamma_1 = 1/6 \approx 0.1666$ for the KPZ class, which is too small compared with the best numerical estimates for this exponent, giving $\alpha_2 \approx 0.388$, as discussed below.

In view of this, let us examine another difference: $\Delta_d = z_d - 2\alpha_d$. Notably, it is simply given by $\Delta_d = d$ for the linear and VLDS equations, so that
\begin{equation}
\Delta_{d+1} - \Delta_d = 1, 
\label{eqDelta}
\end{equation}
for these classes (even if small two-loop corrections are considered in the VLDS case \cite{Janssen}). This demonstrates that the variation $G_{\Delta} \equiv \Delta_{d+1} - \Delta_d$ displays a more universal (and, thus, a simpler) behavior than $G_{\Gamma}$, since $G_{\Delta}$ does not depend neither on $d$ nor on the growth exponents for these classes.

This suggests that $G_{\Delta}$ may behave in a simpler way also for KPZ systems. However, Eq. \ref{eqDelta} does not hold in this case, because $\Delta_2 - \Delta_1 = 1$ furnishes, once again, the incorrect exponent $\alpha_2 = 1/6$. Nevertheless, considering that the KPZ equation may, at least once, present a dimensional variation given by some $\beta$, analogously to $G_{\Gamma}$ for the other classes, then, Eq. \ref{eqAlphaBeta1} is quite suggestive that
\begin{equation}
 \Delta_2 - \Delta_1 = \beta_1
\label{eqDelta2}
\end{equation}
may hold true for the KPZ class. From this relation, and Eqs. \ref{eqBeta} and \ref{eqGI}, one finds the rational exponents
\begin{equation}
\alpha_2 = \frac{7}{18},  \quad\quad \beta_2 = \frac{7}{29} \quad\quad \text{and} \quad\quad z_2 = \frac{29}{18},
\label{eqExp2D}
\end{equation}
giving $\alpha_2 \approx 0.38888$ and $\beta_2 \approx 0.24138$, which are both in striking agreement with the best known numerical estimates for them. For instance, some of the most revered values for the growth exponent were obtained by Kelling \textit{et al.} from very large simulations of an octahedral KPZ model on graphic cards, being $\beta_2 = 0.2415(15)$ \cite{Kelling} and $\beta_2 = 0.2414(2)$ \cite{Kelling18}. Accurate estimates for this exponent were reported also by Halpin-Healy \cite{healy12} --- considering four models for directed polymers in random media (DPRM), the restricted solid-on-solid (RSOS) model \cite{KK} and the numerical integration of the KPZ equation in $d=2$ ---, whose average exponent is again $\beta_2 = 0.2415$. Significantly, this value deviates by less than $0.05$\% from $7/29$, while for the more recent and accurate result $\beta_2 = 0.2414(2)$ \cite{Kelling18} such a difference is only $\approx 0.008$\%. There exist several other less accurate estimates for $\beta_2$ in the literature supporting this rational result; some examples are presented in Tab. \ref{tab1}.

\begin{table*}[t] 
\caption{Numerical estimates for $\beta_d$. The acronyms not defined in the text refer to the ballistic deposition (BD), a directed lattice gas (DLG) of $d$-mers, and `this work' (TW). The average exponents, $\bar{\beta}_d$, are also presented, with the error bars estimated considering an uncertainty `(5)' where it is not available. The analytical results from Eq. \ref{eqExpDqq} are also shown for comparison.}
\begin{ruledtabular}
\centering
\begin{tabular}{llllllllllll}
$d$  & \makecell{HS \\ \cite{Forrest90}} & \makecell{DPRM \\ \cite{HHTake2015}} & \makecell{BD \\ \cite{Alves16}} & \makecell{DLG \\ \cite{Geza10}} & \makecell{RSOS \\ \cite{Alves14}} & \makecell{DPRM \\ \cite{Kim2019,Kim2021}} & \makecell{RSOS \\ \cite{Kim2014}} & \makecell{RSOS \\ {[}TW{]}} & \makecell{DPRM\footnote{Real space RG calculations. The exponents reported in Ref. \cite{Perlsman0} were limited to $d \le 7$, with three decimal digits.} \\ \cite{Perlsman0} / TW} & \makecell{$\bar{\beta}_{d}$} & \makecell{$\dfrac{7}{8d+13}$} \\
\hline\hline
2  & 0.240(1) &           &           & 0.245(5)  &         &           &           &            &  0.2421    & {\bf 0.242(2)}   & 0.241379   \\
3  & 0.180(5) & 0.1868    & 0.185(5)  & 0.184(5)  & 0.190   &           &           &            &  0.1883    & {\bf 0.186(4)}   & 0.189189   \\
4  &          & 0.152(4)  & 0.145(10) & 0.15(1)   & 0.152   &           & 0.158(6)  &            &  0.1539    & {\bf 0.152(6)}   & 0.155556   \\
5  &          &           &           & 0.115(5)  & 0.130   & 0.130(6)  & 0.128(6)  &            &  0.1305    & {\bf 0.127(5)}   & 0.132075   \\
6  &          &           &           &           & 0.105   & 0.110(7)  & 0.108(6)  &            &  0.1135    & {\bf 0.109(5)}   & 0.114754   \\
7  &          &           &           &           &         & 0.102(7)  & 0.096(7)  &            &  0.1007    & {\bf 0.100(5)}   & 0.101449   \\
8  &          &           &           &           &         & 0.092(7)  & 0.086(8)  &            &  0.0906    & {\bf 0.090(5)}   & 0.090909   \\
9  &          &           &           &           &         & 0.078(9)  & 0.077(8)  &            &  0.0825    & {\bf 0.079(6)}   & 0.082353   \\
10 &          &           &           &           &         & 0.069(9)  & 0.070(8)  &            &  0.0758    & {\bf 0.072(6)}   & 0.075269   \\
11 &          &           &           &           &         &           & 0.065(9)  &            &  0.0702    & {\bf 0.068(5)}   & 0.069307   \\
12 &          &           &           &           &         &           &           & 0.061(7)   &  0.0654    & {\bf 0.063(4)}   & 0.064220   \\
13 &          &           &           &           &         &           &           & 0.056(7)   &  0.0613    & {\bf 0.059(4)}   & 0.059829   \\
14 &          &           &           &           &         &           &           & 0.052(7)   &  0.0577    & {\bf 0.055(4)}   & 0.056000   \\
15 &          &           &           &           &         &           &           & 0.049(7)   &  0.0545    & {\bf 0.052(4)}   & 0.052632   \\
\end{tabular}
\end{ruledtabular}
\label{tab1}
\end{table*}

A similar thing happens for the roughness exponent. In fact, considering only the most precise values (with uncertainties at the third or fourth decimal places) found in the literature, one has: $\alpha_2 = 0.385(5)$ for a hypercubic stacking (HS) model \cite{Forrest90}; $\alpha_2 = 0.393(4)$ for an octahedral model \cite{Kelling}; $\alpha_2 = 0.393(3)$ \cite{Marinari} and $\alpha_2 = 0.3869(4)$ \cite{Pagnani} for multi-surface coding of the RSOS model; and $\alpha_2 = 0.387(1)$ for some discrete KPZ models deposited on enlarging substrates \cite{Ismael22}. It is quite remarkable that, considering the error bars, these values either agree with or deviate by less than $0.5$\% from $7/18$. An even more revealing result is the average of these exponents, being $\alpha_2=0.3889$, which differs by only $0.02$\% from $7/18$.

These excellent numerical agreements and the simple properties of $G_{\Gamma}$ and $G_{\Delta}$ for the other universality classes, demonstrating the reliability of Eq. \ref{eqDelta2}, strongly indicate that it is correct/exact. If one assumes that this is true, then, the missing scaling relation for the KPZ class in $d=2$ is simply $\Delta_2 = z_2 - 2\alpha_2 = 5/6$. Note that this relation is obviously not valid for higher dimensions, since the naive generalization $z_d - 2\alpha_d = 5/6$ would give the exponents in Eq. \ref{eqExp2D} for any $d \ge 2$. However, one can find the general behavior of $\Delta_d$ by making two very simple and physically reasonable assumptions: 

\textit{i}) Similarly to the linear and VLDS classes, the KPZ exponents are given by ratios of linear functions of $d$, such that $\Delta_d = (a_1 d + b_1)/(a_2 d + b_2)$, with $a_i,b_i \in \mathbb{Z}$; and

\textit{ii}) $\Delta_d \rightarrow 2$ as $d \rightarrow \infty$, in a way that $a_1/a_2 = 2$. Indeed, at least for $d \rightarrow \infty$, one expects that $\alpha_d \rightarrow 0$ and $z_d \rightarrow 2$ (see, e.g., Ref. \cite{tiago21} for a discussion on this).

\noindent From these two assumptions, along with $\Delta_1=1/2$ and $\Delta_2=5/6$, one readily obtains
\begin{equation}
 z_d - 2\alpha_d = \frac{8 d - 1}{4 d + 10}
\end{equation}
as the missing relation for the KPZ exponents in general. Considering it, together with Eqs. \ref{eqBeta} and \ref{eqGI}, one finds
\begin{equation}
 \alpha_d = \frac{7}{4d+10}, \quad  \beta_d = \frac{7}{8d+13} \quad \text{and} \quad  z_d = \frac{8d+13}{4d+10},
 \label{eqExpDqq}
\end{equation}
which gives the exact exponents for $d=1$ and those in Eq. \ref{eqExp2D} for $d=2$. This is the main result in this work.

To confirm the correctness of these exponents for $d > 2$, let us start taking a close look at the available results for them in the literature. Despite the difficulty in simulating KPZ models for large sizes and long times in high dimensions, robust values for the growth exponent, $\beta_d$, have been reported by various authors, as shown in Tab. \ref{tab1}. There exists also several estimates for the roughness exponent, $\alpha_d$, in the literature, most of them consistent with the values of $\beta_d$ in this table (when compared via Eqs. \ref{eqBeta} and \ref{eqGI}). However, to the best of my knowledge, the available $\alpha_d$'s are limited to $d \le 6$. In fact, the steady state regime is harder to be investigated than the growth regime and, for this reason, I will focus on $\beta_d$ here.

\begin{figure}[b]
 \includegraphics[width=4.cm]{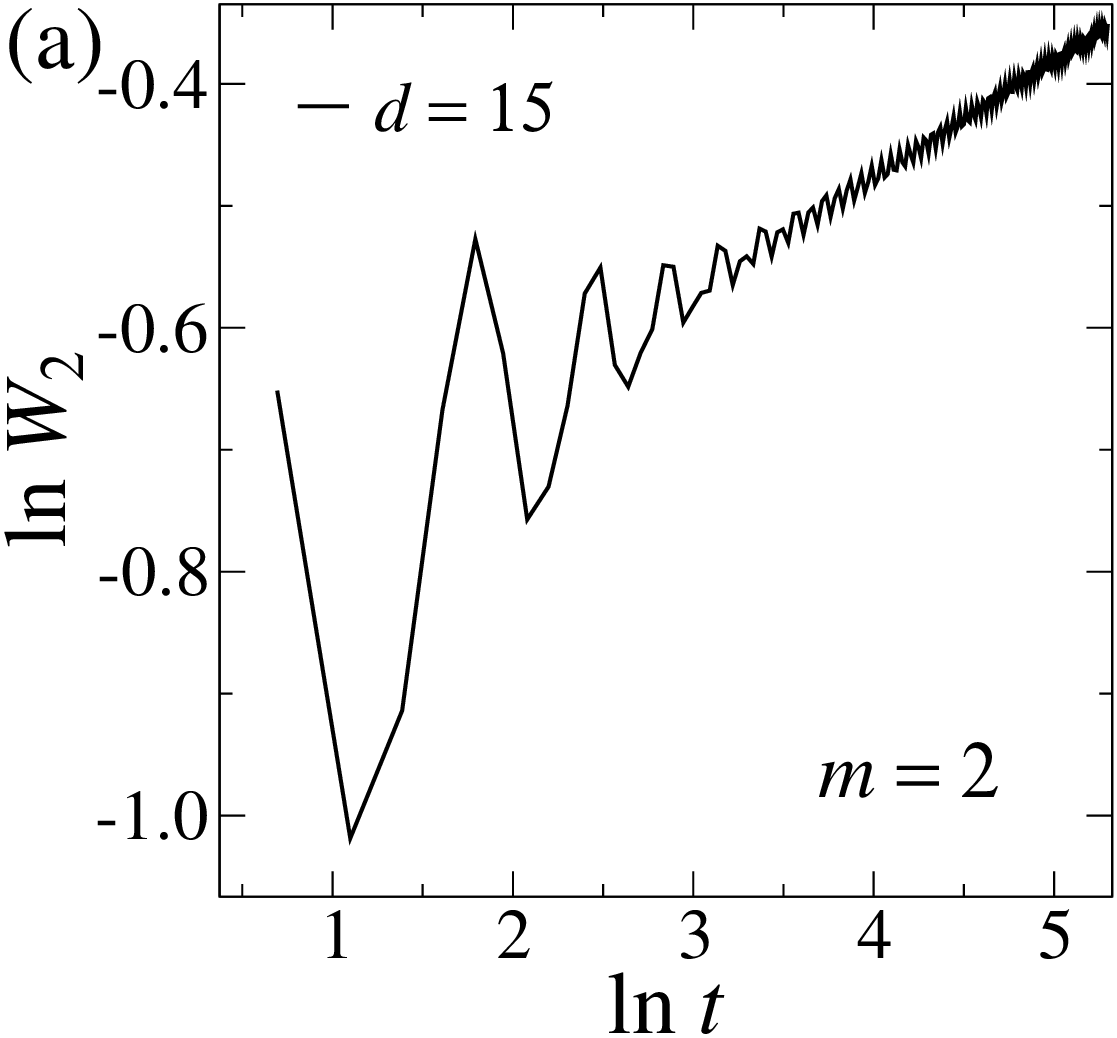}
 \includegraphics[width=4.cm]{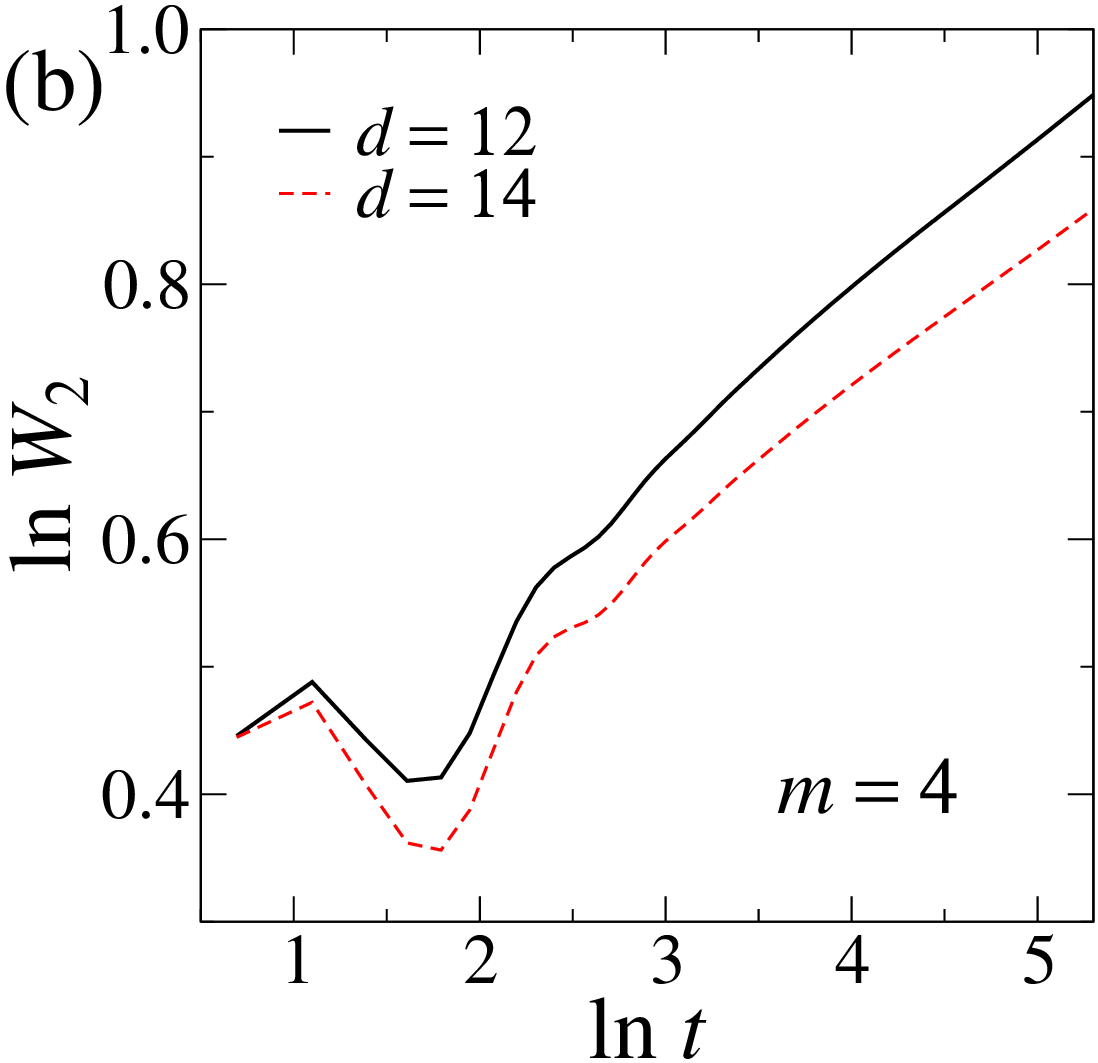}
 \caption{Logarithm of the squared roughness, $W_2$, versus the logarithm of time, $t$, for the generalized RSOS models with a maximum step $m=2$ (a) and $m=4$ (b), for the indicated dimensions.}
\label{figSM1}
\end{figure}

A notable result in Tab. \ref{tab1} is the set of exponents reported by Kim from simulations of DPRM \cite{Kim2021} and RSOS \cite{Kim2014} models, which agree quite well for all dimensions analyzed (up to $d=10$). In order to contribute to the effort of numerically determining the KPZ exponents for very high $d$'s, I performed my own Monte Carlo (MC) simulations of the RSOS models here, extending their results for $d \le 15$. In these models, particles are sequentially and randomly deposited on an initially flat substrate with $N_s$ sites, so that $h_i(t=0)=0$ for $i=1,\ldots, N_s$. Whenever a site $i$ is sorted, a particle aggregates there (i.e., $h_i \rightarrow h_i+1$) if this does not produce a local step larger than a parameter $m$; namely, if $h_i-h_j \le m$, after aggregation, for the $2d$ nearest neighbors $j$ of site $i$. Otherwise, the particle is rejected. The time is updated by $t \rightarrow t + 1/N_s$ after each deposition attempt. I investigate hypercubic substrates with $n$ sides of size $L$ and $(d-n)$ of size $(L-1)$, totaling $N_s=L^n(L-1)^{(d-n)}$ sites \footnote{The ``mixture'' of two lateral sizes ($L$ and $L-1$) is needed in some cases (e.g., $d=14$) because $L^d$ is too large, while $(L-1)^d$ is relatively small.}, considering periodic boundary conditions in all directions. The largest sizes analyzed here were: $L=6$ with $n=9$ for $d=12$, $L=5$ with $n=13$ for $d=13$, $L=5$ with $n=7$ for $d=14$, and $L=4$ with $n=15$ for $d=15$; giving $N_s \gtrsim 10^9$ in all cases. Following Ref. \cite{Kim2014}, I simulate the models for $m=2,3$, and $4$, with 200 samples grown for each one. However, for such large $d$'s I am investigating, the roughness for $m=2$ presents long-lasting oscillations, similar to those observed for $m=1$ in lower dimensions \cite{Kim2014,JMKim13} [see Fig. \ref{figSM1}(a)]. For $m=3$ and $4$, on the other hand, such oscillations disappear at short times, giving rise to a regular scaling regime, as seen in Fig. \ref{figSM1}(b). The average of the effective exponents obtained from these log-log curves of $W_2$ versus $t$ at long times, for $m=3$ and $4$, are shown in Tab. \ref{tab1}.

\begin{figure}[t]
 \includegraphics[width=4.cm]{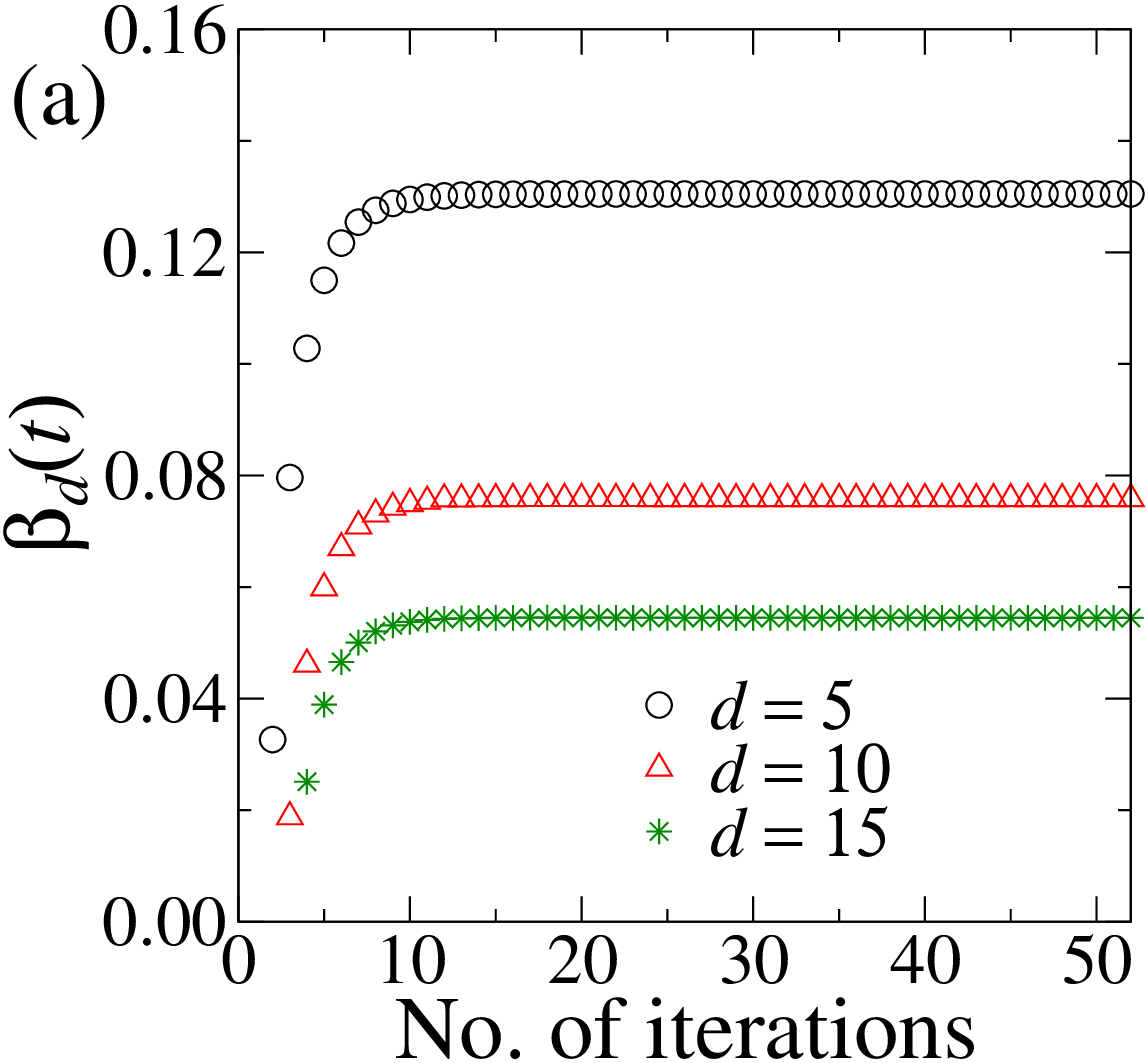}
 \includegraphics[width=4.cm]{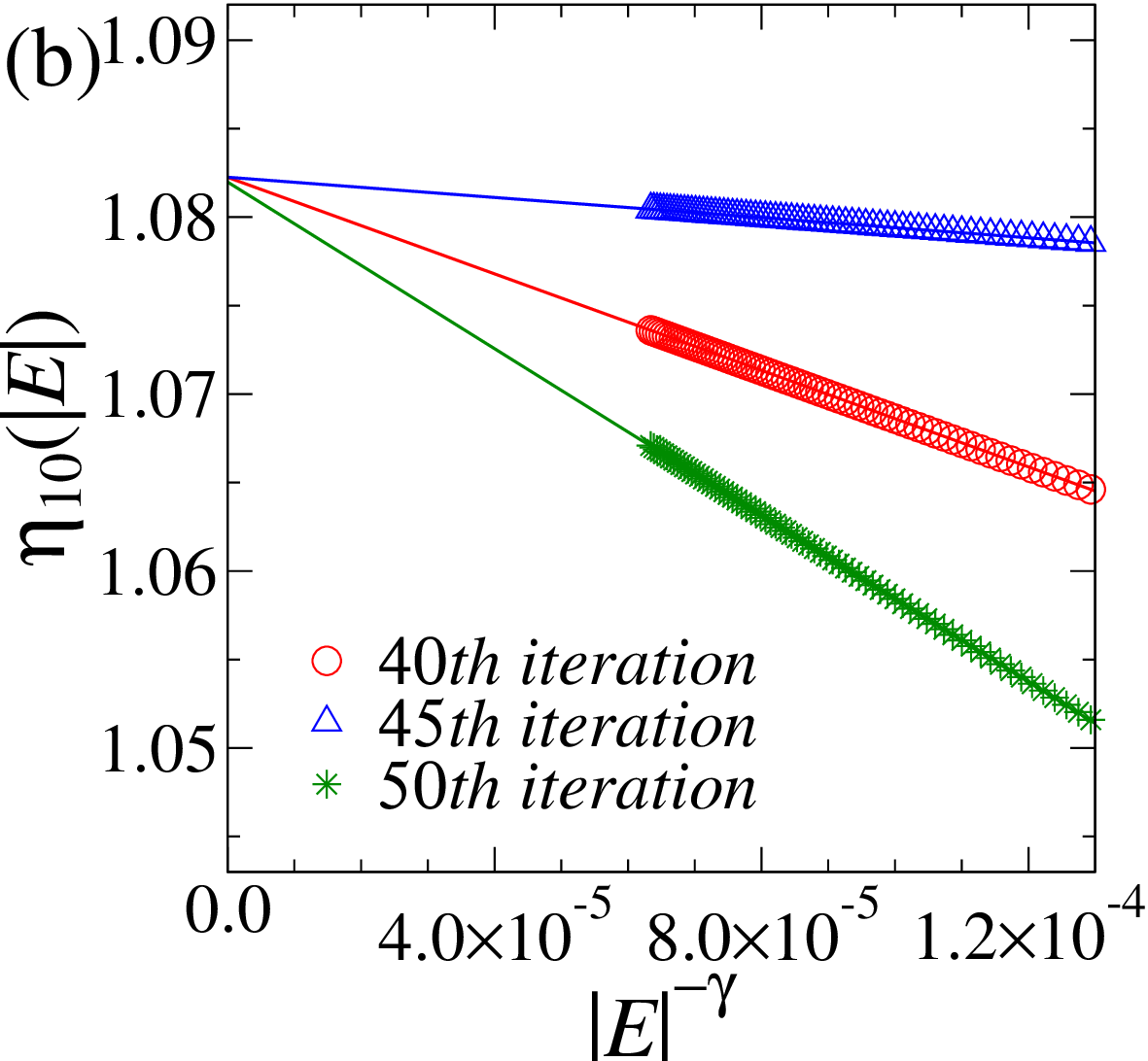}
 \caption{(a) Effective exponents, $\beta_d(t)$, versus the number of iterations of the RG equations (\ref{eqRG1} and \ref{eqRG2}), for the indicated $d$'s. (b) Examples of extrapolations (to the $|E| \rightarrow \infty$ limit) of effective exponents, $\eta_d(|E|)$, for $d=10$. The exponent used in the abscissa to linearize these data was $\gamma=1.05$, and the lines are the linear fits used in the extrapolations.}
\label{figSM2}
\end{figure}

Another remarkable result in Tab. \ref{tab1} is the very good agreement of the exponents obtained by Perlsman and Schwartz (PS) \cite{Perlsman0}, from a real space RG treatment of DPRM, with the rest. Thinking of the $d=1$ case, for example, by considering DPs of length $t$ in a triangular representation --- i.e., on a square lattice rotated by 45{\textdegree} with the origin at its apex ---, the PS approach consists of dividing the central region of the base line into $I_1$ segments of the order of the correlation length ($\sim t^{\nu}$), where $I_1$ is the largest number that still allows to write down the probability $P_t^c(E_{-I_1/2},\ldots,E_{I_1/2})$ of finding a set of paths (in the central region) with ground state energies $\{E_{-I_1/2},\ldots,E_{I_1/2}\}$ as a product of identical and independent PDFs $Q_t(E)$. In this way, PS demonstrated that this PDF is renormalized as
\begin{equation}
 Q_{2t}(E) = I_d \Omega_t(E)\left[ \int_E^{\infty} \Omega_t(\bar{E}) d\bar{E}  \right]^{I_d-1},
\label{eqRG1}
 \end{equation}
where
\begin{equation}
 \Omega_{t}(E) = \int_{-\infty}^{\infty} Q_t(E - \bar{E})Q_t(\bar{E}) d\bar{E}
\label{eqRG2}
\end{equation}
and $I_d= I_1^d$, with $I_1=1.693453$ \cite{Perlsman0}. Then, since the variance, $\sigma^2$, of $Q_t(E)$ is expected to scale asymptotically as $\sigma^2(t) \sim t^{2\beta_d}$, effective growth exponents $\beta_d(t) = \ln[\sigma(2t)/\sigma(t)]/\ln(2)$ obtained by iterating the equations above (starting from a Gaussian with $\sigma = 1$) shall converge to close to the KPZ exponents as $t \rightarrow\infty$. Since these exponents were estimated by PS for $d \le 7$ ``only'' \cite{Perlsman0}, I extended these calculations here for $d\le 15$. Figure \ref{figSM2}(a) shows the behavior of the effective exponents, for some $d$'s, which converge quite fast, yielding accurate estimates for $\beta_d$, as displayed in Tab. \ref{tab1}.

\begin{figure}[t]
 \includegraphics[height=3.7cm]{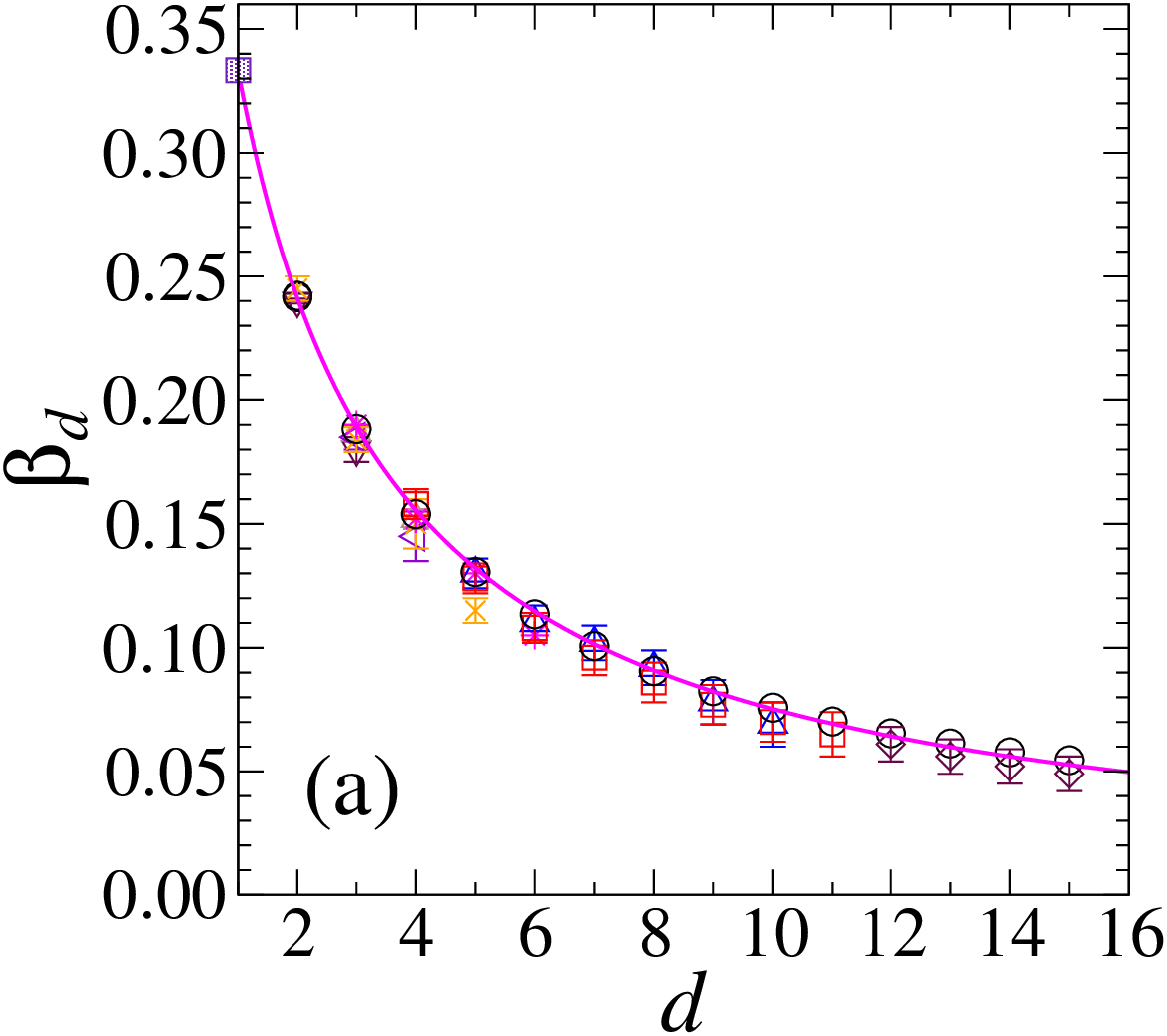}
 \includegraphics[height=3.7cm]{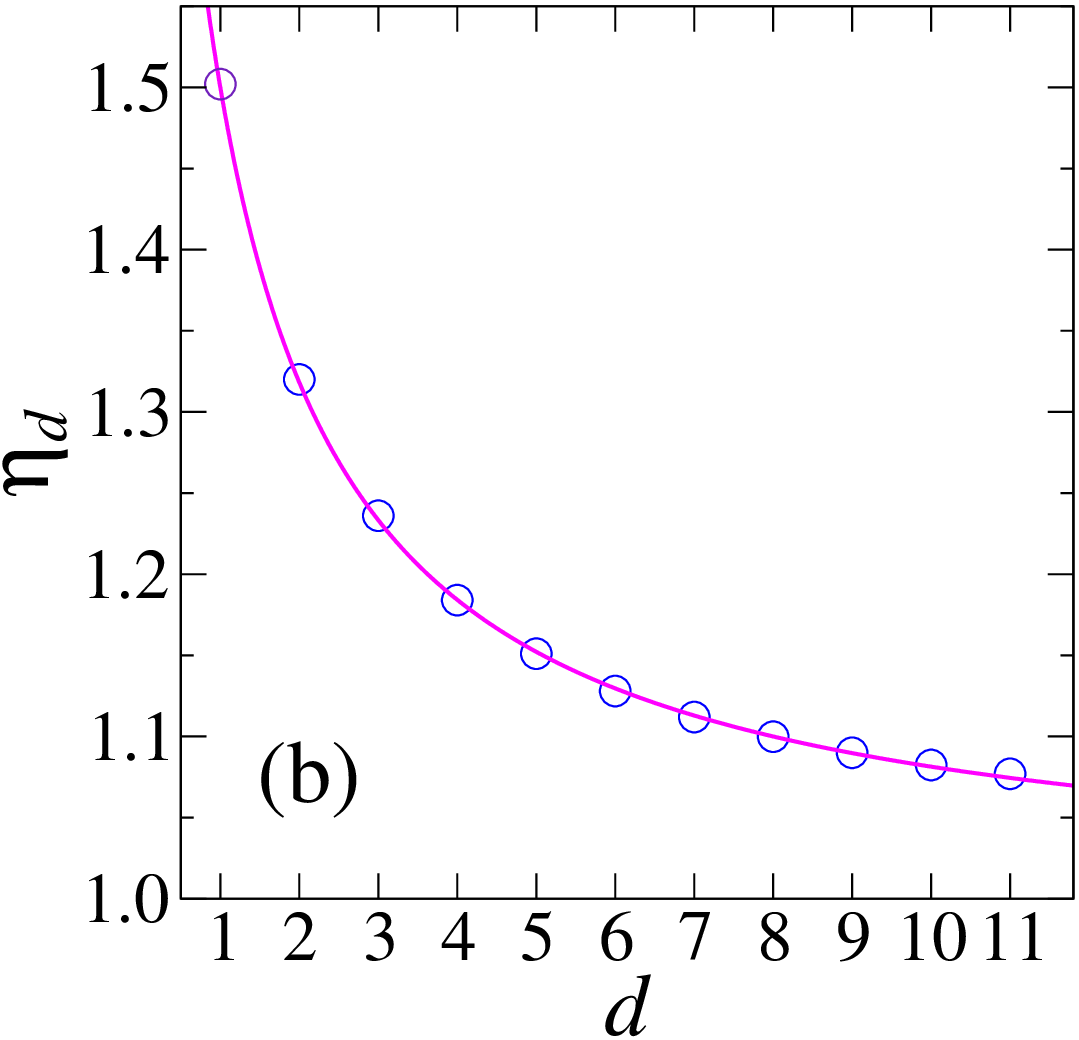}
 \caption{(a) Growth, $\beta_d$, and (b) left-tail, $\eta_d$, exponents as functions of the dimension $d$. The symbols in (a) are the numerical estimates presented in Tab. \ref{tab1} and along the text, while those in (b) are the outcomes from the RG treatment displayed in Tab. \ref{tab2}. The solid lines in (a) and (b) are the analytical results from Eqs. \ref{eqExpDqq} and \ref{eqExpTail}, respectively.}
\label{fig1}
\end{figure}

I remark that such results make it quite clear that the old conjecture by Kim and Kosterlitz \cite{KK} [$\beta_d^{(KK)}=1/(d+2)$], as well as a recent one by Gomes-Filho \textit{et al.} \cite{FernandoExp} [$\beta_d^{(GF)} = (d - \sqrt{(d+1)^2 - 4})/(3-2d)$] are both incorrect. In fact, the results from the former (latter) are systematically larger (smaller) than the numerical estimates in Tab. \ref{tab1}. This is particularly evident for $\beta_d^{(GF)}$, which gives values at least $\sim 1.5\times$ smaller than the average ones (defined as $\bar{\beta}_d$ in the table) for large $d$. On the other hand, the rational exponents found here present a striking agreement with the numerical estimates for all dimensions analyzed, as observed in Tab. \ref{tab1}. For instance, the differences between the analytical and RG values are always smaller than 0.002. Although the exponents from the simulations tend to be slightly smaller, which is certainly a consequence of the very small sizes analyzed, they also agree (considering the uncertainties) with and are very close to $\beta_d=7/(8d+13)$. This is clearly seen in Fig. \ref{fig1}(a), where all these results are compared, strongly indicating that Eq. \ref{eqExpDqq} gives the correct KPZ exponents.

\begin{table*}[t] 
\caption{Numerical estimates for $\eta_d$ from the RG approach compared with the analytical results from Eq. \ref{eqExpTail}.}
\begin{ruledtabular}
\centering
\begin{tabular}{lccccccccccc}
               &   $d=1$    &    2     &    3     &    4     &    5     &    6     &    7     &    8     &    9     &   10     &   11       \\ 
\hline
$\eta_d$ (RG)  &  1.502(2)  & 1.320(2) & 1.236(5) & 1.184(1) & 1.151(1) & 1.128(1) & 1.112(1) & 1.100(1) & 1.090(1) & 1.082(1) & 1.077(3)   \\
$\eta_d$ (Eq. \ref{eqExpTail})  &  1.50000   & 1.31818  & 1.23333  & 1.18421  & 1.15217  & 1.12963  & 1.11290  & 1.10000  & 1.08974  & 1.08140  & 1.07447    \\
\end{tabular}
\end{ruledtabular}
\label{tab2}
\end{table*}

To obtain additional confirmation of this, I investigate also the left tail of the asymptotic PDF $Q(E)$ [i.e.,  $Q_{t \rightarrow \infty}(E)$]. According to the Zhang's argument \cite{healy95}, it behaves as $\ln Q(E) \sim -|E|^{\eta_d}$ as $E \rightarrow -\infty$, with $\eta_d = 1/(1-\beta_d)$ \cite{healy95}. Thereby, from Eq. \ref{eqExpDqq} one gets
\begin{equation}
 \eta_d = \frac{8d+13}{8d+6},
 \label{eqExpTail}
\end{equation}
which gives $\eta_2 = \frac{29}{22} \approx 1.31818$ and $\eta_3 = \frac{37}{30} \approx 1.23333$, in agreement with direct estimates of these exponents from MC simulations of DPs: $1.3 \lesssim \eta_2 \lesssim 1.4$ and $1.15 \lesssim \eta_3 \lesssim 1.25$ \cite{Monthus06}. While it is hard to accurately determine the asymptotic tail behavior from MC simulations, with the RG treatment above $Q_t(E)$ can be obtained for probabilities in a range of hundreds of orders of magnitude. This allows for precise estimates of effective exponents $\eta_d(|E|,t)$, defined as the successive slopes of curves of $\ln[- \ln Q_t(E)]$ versus $\ln |E|$. Extrapolations of these exponents to $|E| \rightarrow \infty$, considering the energy interval $E \in [-10000,-5000]$ and very large $t$'s, yield the results summarized in Tab. \ref{tab2}. [See Fig. \ref{figSM2}(b) for some examples of such extrapolations for $d=10$.] These exponents agree impressively well with Eq. \ref{eqExpTail}, deviating by $\lesssim 0.2$\% from it for all $d$'s analyzed, as it is better appreciated in Fig. \ref{fig1}(b). Besides confirming the correctness of the KPZ exponents in Eq. \ref{eqExpDqq}, which in turn validate Eq. \ref{eqDelta2}, this also demonstrates that the Zhang's formula, $\eta_d = 1/(1-\beta_d)$, holds for high dimensions.

As an aside, I notice that the right-tail exponents, $\eta'_d$, of the DPRM PDFs are harder to be estimated because this tail becomes very steep for large $d$. Nevertheless, I find here that $\eta'_1 \approx 2.64$, $\eta'_2 \approx 3.33$, $\eta'_3 \approx 4.05$ and $\eta'_4 \approx 4.83$ from the RG approach. These values are appreciably smaller than the ones from the Monthus and Garel (MG) \cite{Monthus08} conjecture: $\eta'_d = (d+1)\eta_d$ for DPs on Euclidean lattices. Indeed, it gives $\eta'_1 = 3$, $\eta'_2 \approx 3.9545$, $\eta'_3 \approx 4.9333$ and $\eta'_4 \approx 5.9211$ by using Eq. \ref{eqExpTail} for $\eta_d$. I remark that the RG value above for $d=1$ is the same found in Ref. \cite{HHTake2015} for the DPRM problem on a hierarchical lattice; and, since $\eta'_1=3$ is exact, these discrepancies do not necessarily indicate a failure of the MG conjecture.

In conclusion, the simple assumptions (supported by the dimensional dependence of the exact exponents for other universality classes for interface growth) leading to Eq. \ref{eqExpDqq} (and then to Eq. \ref{eqExpTail}) and the overwhelming numerical confirmation of such predictions, for so many dimensions, strongly indicate that Eq. \ref{eqExpDqq} is the long-awaited exact solution for the KPZ scaling. Since, \textit{a priori}, there exists no reason to expect simple rational numbers for the exact KPZ exponents for all $d$, another possible scenario is that Eq. \ref{eqExpDqq} gives the correct one-loop exponents in exact RG treatments (not developed yet) for the KPZ equation, where two-loop corrections are small, as is the case for the VLDS equation \ref{eqVLDS}. However, the results in Tables \ref{tab1} and \ref{tab2}, as well as in Fig. \ref{fig1} demonstrate that, if some corrections exist to Eq. \ref{eqExpDqq}, they shall be actually almost negligible. Therefore, this provides compelling evidence that the upper critical dimension of the KPZ class is $d_u \rightarrow \infty$, ruling out previous theories yielding a small $d_u$ \cite{Moore,Cates,Bray,Claudin,Healy90,Kinzelbach,Fogedby2}. These findings will certainly motivate and guide future theoretical works intended to rigorously confirm them. In the meantime, for many practical purposes the exponents found here can be used as the KPZ ones, which might be very important, for example, in the study of the KPZ height distributions (HDs) and their geometry dependence in high dimensions. Although these HDs have been numerically characterized for $d=2$ in recent years \cite{healy12,tiago13,healy13,Ismael14,Alves14BD,Ismael22}, with the universality of the PDF for flat geometry being confirmed in some experiments \cite{Almeida14,healy14exp,Almeida15,Almeida17}, for higher $d$'s, only the HDs for the flat case were investigated so far, for $d \le 6$ \cite{Alves14,HHTake2015,Alves16,Kim2019}. With the exponents at hand, these results can be improved and generalized for other geometries and dimensionalities. In case of being exact, these exponents will also pave the way for complete solutions of the KPZ equation in $d>1$, which are particularly relevant for applications in $d=2$.

\acknowledgments

The author acknowledges financial support from CNPq and FAPEMIG (Brazilian agencies); and thanks Fernando A. Oliveira for motivating discussions and Nathann T. Rodrigues for some help with the simulations.

\bibliography{bibExpKPZ2D}

\begin{thebibliography}{63}%
\makeatletter
\providecommand \@ifxundefined [1]{%
 \@ifx{#1\undefined}
}%
\providecommand \@ifnum [1]{%
 \ifnum #1\expandafter \@firstoftwo
 \else \expandafter \@secondoftwo
 \fi
}%
\providecommand \@ifx [1]{%
 \ifx #1\expandafter \@firstoftwo
 \else \expandafter \@secondoftwo
 \fi
}%
\providecommand \natexlab [1]{#1}%
\providecommand \enquote  [1]{``#1''}%
\providecommand \bibnamefont  [1]{#1}%
\providecommand \bibfnamefont [1]{#1}%
\providecommand \citenamefont [1]{#1}%
\providecommand \href@noop [0]{\@secondoftwo}%
\providecommand \href [0]{\begingroup \@sanitize@url \@href}%
\providecommand \@href[1]{\@@startlink{#1}\@@href}%
\providecommand \@@href[1]{\endgroup#1\@@endlink}%
\providecommand \@sanitize@url [0]{\catcode `\\12\catcode `\$12\catcode
  `\&12\catcode `\#12\catcode `\^12\catcode `\_12\catcode `\%12\relax}%
\providecommand \@@startlink[1]{}%
\providecommand \@@endlink[0]{}%
\providecommand \url  [0]{\begingroup\@sanitize@url \@url }%
\providecommand \@url [1]{\endgroup\@href {#1}{\urlprefix }}%
\providecommand \urlprefix  [0]{URL }%
\providecommand \Eprint [0]{\href }%
\providecommand \doibase [0]{http://dx.doi.org/}%
\providecommand \selectlanguage [0]{\@gobble}%
\providecommand \bibinfo  [0]{\@secondoftwo}%
\providecommand \bibfield  [0]{\@secondoftwo}%
\providecommand \translation [1]{[#1]}%
\providecommand \BibitemOpen [0]{}%
\providecommand \bibitemStop [0]{}%
\providecommand \bibitemNoStop [0]{.\EOS\space}%
\providecommand \EOS [0]{\spacefactor3000\relax}%
\providecommand \BibitemShut  [1]{\csname bibitem#1\endcsname}%
\let\auto@bib@innerbib\@empty
%</preamble>
\bibitem [{\citenamefont {Kardar}\ \emph {et~al.}(1986)\citenamefont {Kardar},
  \citenamefont {Parisi},\ and\ \citenamefont {Zhang}}]{KPZ}%
  \BibitemOpen
  \bibfield  {author} {\bibinfo {author} {\bibfnamefont {M.}~\bibnamefont
  {Kardar}}, \bibinfo {author} {\bibfnamefont {G.}~\bibnamefont {Parisi}}, \
  and\ \bibinfo {author} {\bibfnamefont {Y.-C.}\ \bibnamefont {Zhang}},\
  }\href@noop {} {\bibfield  {journal} {\bibinfo  {journal} {Phys. Rev. Lett.}\
  }\textbf {\bibinfo {volume} {56}},\ \bibinfo {pages} {889} (\bibinfo {year}
  {1986})}\BibitemShut {NoStop}%
\bibitem [{\citenamefont {Barabasi}\ and\ \citenamefont
  {Stanley}(1995)}]{barabasi}%
  \BibitemOpen
  \bibfield  {author} {\bibinfo {author} {\bibfnamefont {A.-L.}\ \bibnamefont
  {Barabasi}}\ and\ \bibinfo {author} {\bibfnamefont {H.~E.}\ \bibnamefont
  {Stanley}},\ }\href@noop {} {\emph {\bibinfo {title} {{Fractal Concepts in
  Surface Growth}}}}\ (\bibinfo  {publisher} {Cambridge University Press},\
  \bibinfo {address} {Cambridge, England},\ \bibinfo {year} {1995})\BibitemShut
  {NoStop}%
\bibitem [{\citenamefont {Halpin-Healy}\ and\ \citenamefont
  {Zhang}(1995)}]{healy95}%
  \BibitemOpen
  \bibfield  {author} {\bibinfo {author} {\bibfnamefont {T.}~\bibnamefont
  {Halpin-Healy}}\ and\ \bibinfo {author} {\bibfnamefont {Y.~C.}\ \bibnamefont
  {Zhang}},\ }\href {\doibase 10.1016/0370-1573(94)00087-J} {\bibfield
  {journal} {\bibinfo  {journal} {Phys. Rep.}\ }\textbf {\bibinfo {volume}
  {254}},\ \bibinfo {pages} {215} (\bibinfo {year} {1995})}\BibitemShut
  {NoStop}%
\bibitem [{\citenamefont {Krug}(1997)}]{KrugAdv}%
  \BibitemOpen
  \bibfield  {author} {\bibinfo {author} {\bibfnamefont {J.}~\bibnamefont
  {Krug}},\ }\href@noop {} {\bibfield  {journal} {\bibinfo  {journal} {Adv.
  Phys.}\ }\textbf {\bibinfo {volume} {46}},\ \bibinfo {pages} {139} (\bibinfo
  {year} {1997})}\BibitemShut {NoStop}%
\bibitem [{\citenamefont {Kriecherbauer}\ and\ \citenamefont
  {Krug}(2010)}]{Kriecherbauer2010}%
  \BibitemOpen
  \bibfield  {author} {\bibinfo {author} {\bibfnamefont {T.}~\bibnamefont
  {Kriecherbauer}}\ and\ \bibinfo {author} {\bibfnamefont {J.}~\bibnamefont
  {Krug}},\ }\href {\doibase 10.1088/1751-8113/43/40/403001} {\bibfield
  {journal} {\bibinfo  {journal} {J. Phys. A Math. Theor.}\ }\textbf {\bibinfo
  {volume} {43}},\ \bibinfo {pages} {403001} (\bibinfo {year}
  {2010})}\BibitemShut {NoStop}%
\bibitem [{\citenamefont {Brush}(1967)}]{Brush67}%
  \BibitemOpen
  \bibfield  {author} {\bibinfo {author} {\bibfnamefont {S.~G.}\ \bibnamefont
  {Brush}},\ }\href@noop {} {\bibfield  {journal} {\bibinfo  {journal} {Rev.
  Mod. Phys.}\ }\textbf {\bibinfo {volume} {39}},\ \bibinfo {pages} {883}
  (\bibinfo {year} {1967})}\BibitemShut {NoStop}%
\bibitem [{\citenamefont {Sasamoto}\ and\ \citenamefont
  {Spohn}(2010)}]{Sasamoto2010}%
  \BibitemOpen
  \bibfield  {author} {\bibinfo {author} {\bibfnamefont {T.}~\bibnamefont
  {Sasamoto}}\ and\ \bibinfo {author} {\bibfnamefont {H.}~\bibnamefont
  {Spohn}},\ }\href@noop {} {\bibfield  {journal} {\bibinfo  {journal} {Phys.
  Rev. Lett.}\ }\textbf {\bibinfo {volume} {104}},\ \bibinfo {pages} {1}
  (\bibinfo {year} {2010})}\BibitemShut {NoStop}%
\bibitem [{\citenamefont {Amir}\ \emph {et~al.}(2011)\citenamefont {Amir},
  \citenamefont {Corwin},\ and\ \citenamefont {Quastel}}]{Amir}%
  \BibitemOpen
  \bibfield  {author} {\bibinfo {author} {\bibfnamefont {G.}~\bibnamefont
  {Amir}}, \bibinfo {author} {\bibfnamefont {I.}~\bibnamefont {Corwin}}, \ and\
  \bibinfo {author} {\bibfnamefont {J.}~\bibnamefont {Quastel}},\ }\href@noop
  {} {\bibfield  {journal} {\bibinfo  {journal} {Commun. Pure Appl. Math.}\
  }\textbf {\bibinfo {volume} {64}},\ \bibinfo {pages} {466} (\bibinfo {year}
  {2011})}\BibitemShut {NoStop}%
\bibitem [{\citenamefont {Cardy}(1996)}]{Cardy}%
  \BibitemOpen
  \bibfield  {author} {\bibinfo {author} {\bibfnamefont {J.}~\bibnamefont
  {Cardy}},\ }\href {\doibase 10.1017/CBO9781316036440} {\emph {\bibinfo
  {title} {{Scaling and Renormalization in Statistical Physics}}}}\ (\bibinfo
  {publisher} {Cambridge University Press},\ \bibinfo {address} {Cambridge,
  England},\ \bibinfo {year} {1996})\BibitemShut {NoStop}%
\bibitem [{\citenamefont {Colaiori}\ and\ \citenamefont {Moore}(2001)}]{Moore}%
  \BibitemOpen
  \bibfield  {author} {\bibinfo {author} {\bibfnamefont {F.}~\bibnamefont
  {Colaiori}}\ and\ \bibinfo {author} {\bibfnamefont {M.~A.}\ \bibnamefont
  {Moore}},\ }\href {\doibase 10.1103/PhysRevLett.86.3946} {\bibfield
  {journal} {\bibinfo  {journal} {Phys. Rev. Lett.}\ }\textbf {\bibinfo
  {volume} {86}},\ \bibinfo {pages} {3946} (\bibinfo {year}
  {2001})}\BibitemShut {NoStop}%
\bibitem [{\citenamefont {Bouchaud}\ and\ \citenamefont {Cates}(1993)}]{Cates}%
  \BibitemOpen
  \bibfield  {author} {\bibinfo {author} {\bibfnamefont {J.~P.}\ \bibnamefont
  {Bouchaud}}\ and\ \bibinfo {author} {\bibfnamefont {M.~E.}\ \bibnamefont
  {Cates}},\ }\href@noop {} {\bibfield  {journal} {\bibinfo  {journal} {Phys.
  Rev. E}\ }\textbf {\bibinfo {volume} {47}},\ \bibinfo {pages} {R1455}
  (\bibinfo {year} {1993})}\BibitemShut {NoStop}%
\bibitem [{\citenamefont {Doherty}\ \emph {et~al.}(1994)\citenamefont
  {Doherty}, \citenamefont {Moore}, \citenamefont {Kim},\ and\ \citenamefont
  {Bray}}]{Bray}%
  \BibitemOpen
  \bibfield  {author} {\bibinfo {author} {\bibfnamefont {J.~P.}\ \bibnamefont
  {Doherty}}, \bibinfo {author} {\bibfnamefont {M.~A.}\ \bibnamefont {Moore}},
  \bibinfo {author} {\bibfnamefont {J.~M.}\ \bibnamefont {Kim}}, \ and\
  \bibinfo {author} {\bibfnamefont {A.~J.}\ \bibnamefont {Bray}},\ }\href@noop
  {} {\bibfield  {journal} {\bibinfo  {journal} {Phys. Rev. Lett.}\ }\textbf
  {\bibinfo {volume} {72}},\ \bibinfo {pages} {2041} (\bibinfo {year}
  {1994})}\BibitemShut {NoStop}%
\bibitem [{\citenamefont {Moore}\ \emph {et~al.}(1995)\citenamefont {Moore},
  \citenamefont {Blum}, \citenamefont {Doherty}, \citenamefont {Marsili},
  \citenamefont {Bouchaud},\ and\ \citenamefont {Claudin}}]{Claudin}%
  \BibitemOpen
  \bibfield  {author} {\bibinfo {author} {\bibfnamefont {M.~A.}\ \bibnamefont
  {Moore}}, \bibinfo {author} {\bibfnamefont {T.}~\bibnamefont {Blum}},
  \bibinfo {author} {\bibfnamefont {J.~P.}\ \bibnamefont {Doherty}}, \bibinfo
  {author} {\bibfnamefont {M.}~\bibnamefont {Marsili}}, \bibinfo {author}
  {\bibfnamefont {J.-P.}\ \bibnamefont {Bouchaud}}, \ and\ \bibinfo {author}
  {\bibfnamefont {P.}~\bibnamefont {Claudin}},\ }\href@noop {} {\bibfield
  {journal} {\bibinfo  {journal} {Phys. Rev. Lett.}\ }\textbf {\bibinfo
  {volume} {74}},\ \bibinfo {pages} {4257} (\bibinfo {year}
  {1995})}\BibitemShut {NoStop}%
\bibitem [{\citenamefont {{Halpin-Healy}}(1990)}]{Healy90}%
  \BibitemOpen
  \bibfield  {author} {\bibinfo {author} {\bibfnamefont {T.}~\bibnamefont
  {{Halpin-Healy}}},\ }\href@noop {} {\bibfield  {journal} {\bibinfo  {journal}
  {Phys. Rev. A}\ }\textbf {\bibinfo {volume} {42}},\ \bibinfo {pages} {711}
  (\bibinfo {year} {1990})}\BibitemShut {NoStop}%
\bibitem [{\citenamefont {L\"assig}\ and\ \citenamefont
  {Kinzelbach}(1997)}]{Kinzelbach}%
  \BibitemOpen
  \bibfield  {author} {\bibinfo {author} {\bibfnamefont {M.}~\bibnamefont
  {L\"assig}}\ and\ \bibinfo {author} {\bibfnamefont {H.}~\bibnamefont
  {Kinzelbach}},\ }\href@noop {} {\bibfield  {journal} {\bibinfo  {journal}
  {Phys. Rev. Lett.}\ }\textbf {\bibinfo {volume} {78}},\ \bibinfo {pages}
  {903} (\bibinfo {year} {1997})}\BibitemShut {NoStop}%
\bibitem [{\citenamefont {Fogedby}(2006)}]{Fogedby2}%
  \BibitemOpen
  \bibfield  {author} {\bibinfo {author} {\bibfnamefont {H.~C.}\ \bibnamefont
  {Fogedby}},\ }\href@noop {} {\bibfield  {journal} {\bibinfo  {journal} {Phys.
  Rev. E}\ }\textbf {\bibinfo {volume} {73}},\ \bibinfo {pages} {031104}
  (\bibinfo {year} {2006})}\BibitemShut {NoStop}%
\bibitem [{\citenamefont {Kloss}\ \emph
  {et~al.}(2014{\natexlab{a}})\citenamefont {Kloss}, \citenamefont {Canet},
  \citenamefont {Delamotte},\ and\ \citenamefont {Wschebor}}]{Canet}%
  \BibitemOpen
  \bibfield  {author} {\bibinfo {author} {\bibfnamefont {T.}~\bibnamefont
  {Kloss}}, \bibinfo {author} {\bibfnamefont {L.}~\bibnamefont {Canet}},
  \bibinfo {author} {\bibfnamefont {B.}~\bibnamefont {Delamotte}}, \ and\
  \bibinfo {author} {\bibfnamefont {N.}~\bibnamefont {Wschebor}},\ }\href@noop
  {} {\bibfield  {journal} {\bibinfo  {journal} {Phys. Rev. E}\ }\textbf
  {\bibinfo {volume} {89}},\ \bibinfo {pages} {022108} (\bibinfo {year}
  {2014}{\natexlab{a}})}\BibitemShut {NoStop}%
\bibitem [{\citenamefont {Kloss}\ \emph
  {et~al.}(2014{\natexlab{b}})\citenamefont {Kloss}, \citenamefont {Canet},\
  and\ \citenamefont {Wschebor}}]{Kloss}%
  \BibitemOpen
  \bibfield  {author} {\bibinfo {author} {\bibfnamefont {T.}~\bibnamefont
  {Kloss}}, \bibinfo {author} {\bibfnamefont {L.}~\bibnamefont {Canet}}, \ and\
  \bibinfo {author} {\bibfnamefont {N.}~\bibnamefont {Wschebor}},\ }\href@noop
  {} {\bibfield  {journal} {\bibinfo  {journal} {Phys. Rev. E}\ }\textbf
  {\bibinfo {volume} {90}},\ \bibinfo {pages} {062133} (\bibinfo {year}
  {2014}{\natexlab{b}})}\BibitemShut {NoStop}%
\bibitem [{\citenamefont {Perlsman}\ and\ \citenamefont
  {Schwartz}(1996)}]{Perlsman0}%
  \BibitemOpen
  \bibfield  {author} {\bibinfo {author} {\bibfnamefont {E.}~\bibnamefont
  {Perlsman}}\ and\ \bibinfo {author} {\bibfnamefont {M.}~\bibnamefont
  {Schwartz}},\ }\href@noop {} {\bibfield  {journal} {\bibinfo  {journal}
  {Physica A}\ }\textbf {\bibinfo {volume} {234}},\ \bibinfo {pages} {523}
  (\bibinfo {year} {1996})}\BibitemShut {NoStop}%
\bibitem [{\citenamefont {Castellano}\ \emph
  {et~al.}(1998{\natexlab{a}})\citenamefont {Castellano}, \citenamefont
  {Marsili},\ and\ \citenamefont {Pietronero}}]{Castellano1}%
  \BibitemOpen
  \bibfield  {author} {\bibinfo {author} {\bibfnamefont {C.}~\bibnamefont
  {Castellano}}, \bibinfo {author} {\bibfnamefont {M.}~\bibnamefont {Marsili}},
  \ and\ \bibinfo {author} {\bibfnamefont {L.}~\bibnamefont {Pietronero}},\
  }\href@noop {} {\bibfield  {journal} {\bibinfo  {journal} {Phys. Rev. Lett.}\
  }\textbf {\bibinfo {volume} {80}},\ \bibinfo {pages} {3527} (\bibinfo {year}
  {1998}{\natexlab{a}})}\BibitemShut {NoStop}%
\bibitem [{\citenamefont {Castellano}\ \emph
  {et~al.}(1998{\natexlab{b}})\citenamefont {Castellano}, \citenamefont
  {Gabrielli}, \citenamefont {Marsili}, \citenamefont {Munoz},\ and\
  \citenamefont {Pietronero}}]{Castellano2}%
  \BibitemOpen
  \bibfield  {author} {\bibinfo {author} {\bibfnamefont {C.}~\bibnamefont
  {Castellano}}, \bibinfo {author} {\bibfnamefont {A.}~\bibnamefont
  {Gabrielli}}, \bibinfo {author} {\bibfnamefont {M.}~\bibnamefont {Marsili}},
  \bibinfo {author} {\bibfnamefont {M.~A.}\ \bibnamefont {Munoz}}, \ and\
  \bibinfo {author} {\bibfnamefont {L.}~\bibnamefont {Pietronero}},\
  }\href@noop {} {\bibfield  {journal} {\bibinfo  {journal} {Phys. Rev. E}\
  }\textbf {\bibinfo {volume} {58}},\ \bibinfo {pages} {R5209} (\bibinfo {year}
  {1998}{\natexlab{b}})}\BibitemShut {NoStop}%
\bibitem [{\citenamefont {Ala-Nissila}\ \emph {et~al.}(1993)\citenamefont
  {Ala-Nissila}, \citenamefont {Hjelt}, \citenamefont {Kosterlitz},\ and\
  \citenamefont {Ven\"al\"ainen}}]{Ala1}%
  \BibitemOpen
  \bibfield  {author} {\bibinfo {author} {\bibfnamefont {T.}~\bibnamefont
  {Ala-Nissila}}, \bibinfo {author} {\bibfnamefont {T.}~\bibnamefont {Hjelt}},
  \bibinfo {author} {\bibfnamefont {J.}~\bibnamefont {Kosterlitz}}, \ and\
  \bibinfo {author} {\bibfnamefont {O.}~\bibnamefont {Ven\"al\"ainen}},\
  }\href@noop {} {\bibfield  {journal} {\bibinfo  {journal} {J. Stat. Phys.}\
  }\textbf {\bibinfo {volume} {72}},\ \bibinfo {pages} {207} (\bibinfo {year}
  {1993})}\BibitemShut {NoStop}%
\bibitem [{\citenamefont {Ala-Nissila}(1998)}]{Ala2}%
  \BibitemOpen
  \bibfield  {author} {\bibinfo {author} {\bibfnamefont {T.}~\bibnamefont
  {Ala-Nissila}},\ }\href@noop {} {\bibfield  {journal} {\bibinfo  {journal}
  {Phys. Rev. Lett.}\ }\textbf {\bibinfo {volume} {80}},\ \bibinfo {pages}
  {887} (\bibinfo {year} {1998})}\BibitemShut {NoStop}%
\bibitem [{\citenamefont {Marinari}\ \emph {et~al.}(2002)\citenamefont
  {Marinari}, \citenamefont {Pagnani}, \citenamefont {Parisi},\ and\
  \citenamefont {R\'acz}}]{Marinari02}%
  \BibitemOpen
  \bibfield  {author} {\bibinfo {author} {\bibfnamefont {E.}~\bibnamefont
  {Marinari}}, \bibinfo {author} {\bibfnamefont {A.}~\bibnamefont {Pagnani}},
  \bibinfo {author} {\bibfnamefont {G.}~\bibnamefont {Parisi}}, \ and\ \bibinfo
  {author} {\bibfnamefont {Z.}~\bibnamefont {R\'acz}},\ }\href@noop {}
  {\bibfield  {journal} {\bibinfo  {journal} {Phys. Rev. E}\ }\textbf {\bibinfo
  {volume} {65}},\ \bibinfo {pages} {026136} (\bibinfo {year}
  {2002})}\BibitemShut {NoStop}%
\bibitem [{\citenamefont {Perlsman}\ and\ \citenamefont
  {Havlin}(2006)}]{Perlsman2}%
  \BibitemOpen
  \bibfield  {author} {\bibinfo {author} {\bibfnamefont {E.}~\bibnamefont
  {Perlsman}}\ and\ \bibinfo {author} {\bibfnamefont {S.}~\bibnamefont
  {Havlin}},\ }\href@noop {} {\bibfield  {journal} {\bibinfo  {journal}
  {Europhys. Lett.}\ }\textbf {\bibinfo {volume} {73}},\ \bibinfo {pages} {178}
  (\bibinfo {year} {2006})}\BibitemShut {NoStop}%
\bibitem [{\citenamefont {Schwartz}\ and\ \citenamefont
  {Perlsman}(2012)}]{Perlsman1}%
  \BibitemOpen
  \bibfield  {author} {\bibinfo {author} {\bibfnamefont {M.}~\bibnamefont
  {Schwartz}}\ and\ \bibinfo {author} {\bibfnamefont {E.}~\bibnamefont
  {Perlsman}},\ }\href@noop {} {\bibfield  {journal} {\bibinfo  {journal}
  {Phys. Rev. E}\ }\textbf {\bibinfo {volume} {85}},\ \bibinfo {pages} {050103}
  (\bibinfo {year} {2012})}\BibitemShut {NoStop}%
\bibitem [{\citenamefont {\'Odor}\ \emph {et~al.}(2010)\citenamefont {\'Odor},
  \citenamefont {Liedke},\ and\ \citenamefont {Heinig}}]{Geza10}%
  \BibitemOpen
  \bibfield  {author} {\bibinfo {author} {\bibfnamefont {G.}~\bibnamefont
  {\'Odor}}, \bibinfo {author} {\bibfnamefont {B.}~\bibnamefont {Liedke}}, \
  and\ \bibinfo {author} {\bibfnamefont {K.-H.}\ \bibnamefont {Heinig}},\
  }\href {\doibase 10.1103/PhysRevE.81.031112} {\bibfield  {journal} {\bibinfo
  {journal} {Phys. Rev. E}\ }\textbf {\bibinfo {volume} {81}},\ \bibinfo
  {pages} {031112} (\bibinfo {year} {2010})}\BibitemShut {NoStop}%
\bibitem [{\citenamefont {Pagnani}\ and\ \citenamefont
  {Parisi}(2013)}]{Parisi13}%
  \BibitemOpen
  \bibfield  {author} {\bibinfo {author} {\bibfnamefont {A.}~\bibnamefont
  {Pagnani}}\ and\ \bibinfo {author} {\bibfnamefont {G.}~\bibnamefont
  {Parisi}},\ }\href@noop {} {\bibfield  {journal} {\bibinfo  {journal} {Phys.
  Rev. E}\ }\textbf {\bibinfo {volume} {87}},\ \bibinfo {pages} {010102}
  (\bibinfo {year} {2013})}\BibitemShut {NoStop}%
\bibitem [{\citenamefont {Rodrigues}\ \emph {et~al.}(2014)\citenamefont
  {Rodrigues}, \citenamefont {Mello},\ and\ \citenamefont
  {Oliveira}}]{Fernando}%
  \BibitemOpen
  \bibfield  {author} {\bibinfo {author} {\bibfnamefont {E.~A.}\ \bibnamefont
  {Rodrigues}}, \bibinfo {author} {\bibfnamefont {B.~A.}\ \bibnamefont
  {Mello}}, \ and\ \bibinfo {author} {\bibfnamefont {F.~A.}\ \bibnamefont
  {Oliveira}},\ }\href {\doibase 10.1088/1751-8113/48/3/035001} {\bibfield
  {journal} {\bibinfo  {journal} {J. Phys. A: Math. Theor.}\ }\textbf {\bibinfo
  {volume} {48}},\ \bibinfo {pages} {035001} (\bibinfo {year}
  {2014})}\BibitemShut {NoStop}%
\bibitem [{\citenamefont {Kim}\ and\ \citenamefont {Kim}(2013)}]{JMKim13}%
  \BibitemOpen
  \bibfield  {author} {\bibinfo {author} {\bibfnamefont {J.~M.}\ \bibnamefont
  {Kim}}\ and\ \bibinfo {author} {\bibfnamefont {S.-W.}\ \bibnamefont {Kim}},\
  }\href {\doibase 10.1103/PhysRevE.88.034102} {\bibfield  {journal} {\bibinfo
  {journal} {Phys. Rev. E}\ }\textbf {\bibinfo {volume} {88}},\ \bibinfo
  {pages} {034102} (\bibinfo {year} {2013})}\BibitemShut {NoStop}%
\bibitem [{\citenamefont {Alves}\ \emph
  {et~al.}(2014{\natexlab{a}})\citenamefont {Alves}, \citenamefont {Oliveira},\
  and\ \citenamefont {Ferreira}}]{Alves14}%
  \BibitemOpen
  \bibfield  {author} {\bibinfo {author} {\bibfnamefont {S.~G.}\ \bibnamefont
  {Alves}}, \bibinfo {author} {\bibfnamefont {T.~J.}\ \bibnamefont {Oliveira}},
  \ and\ \bibinfo {author} {\bibfnamefont {S.~C.}\ \bibnamefont {Ferreira}},\
  }\href {\doibase 10.1103/PhysRevE.90.020103} {\bibfield  {journal} {\bibinfo
  {journal} {Phys. Rev. E}\ }\textbf {\bibinfo {volume} {90}},\ \bibinfo
  {pages} {020103(R)} (\bibinfo {year} {2014}{\natexlab{a}})}\BibitemShut
  {NoStop}%
\bibitem [{\citenamefont {Kim}(2019)}]{Kim2019}%
  \BibitemOpen
  \bibfield  {author} {\bibinfo {author} {\bibfnamefont {J.~M.}\ \bibnamefont
  {Kim}},\ }\href {\doibase 10.1088/1742-5468/ab4e8c} {\bibfield  {journal}
  {\bibinfo  {journal} {J. Stat. Mech.}\ }\textbf {\bibinfo {volume} {2019}},\
  \bibinfo {pages} {123206} (\bibinfo {year} {2019})}\BibitemShut {NoStop}%
\bibitem [{\citenamefont {Halpin-Healy}\ and\ \citenamefont
  {Takeuchi}(2015)}]{HHTake2015}%
  \BibitemOpen
  \bibfield  {author} {\bibinfo {author} {\bibfnamefont {T.}~\bibnamefont
  {Halpin-Healy}}\ and\ \bibinfo {author} {\bibfnamefont {K.~A.}\ \bibnamefont
  {Takeuchi}},\ }\href {\doibase 10.1007/s10955-015-1282-1} {\bibfield
  {journal} {\bibinfo  {journal} {J. Stat. Phys.}\ }\textbf {\bibinfo {volume}
  {160}},\ \bibinfo {pages} {794} (\bibinfo {year} {2015})}\BibitemShut
  {NoStop}%
\bibitem [{\citenamefont {Alves}\ and\ \citenamefont
  {Ferreira}(2016)}]{Alves16}%
  \BibitemOpen
  \bibfield  {author} {\bibinfo {author} {\bibfnamefont {S.~G.}\ \bibnamefont
  {Alves}}\ and\ \bibinfo {author} {\bibfnamefont {S.~C.}\ \bibnamefont
  {Ferreira}},\ }\href {\doibase 10.1103/PhysRevE.93.052131} {\bibfield
  {journal} {\bibinfo  {journal} {Phys. Rev. E}\ }\textbf {\bibinfo {volume}
  {93}},\ \bibinfo {pages} {052131} (\bibinfo {year} {2016})}\BibitemShut
  {NoStop}%
\bibitem [{\citenamefont {L\"assig}(1998)}]{Lassig}%
  \BibitemOpen
  \bibfield  {author} {\bibinfo {author} {\bibfnamefont {M.}~\bibnamefont
  {L\"assig}},\ }\href {\doibase 10.1103/PhysRevLett.80.2366} {\bibfield
  {journal} {\bibinfo  {journal} {Phys. Rev. Lett.}\ }\textbf {\bibinfo
  {volume} {80}},\ \bibinfo {pages} {2366} (\bibinfo {year}
  {1998})}\BibitemShut {NoStop}%
\bibitem [{\citenamefont {Fogedby}(2005)}]{Fogedby}%
  \BibitemOpen
  \bibfield  {author} {\bibinfo {author} {\bibfnamefont {H.~C.}\ \bibnamefont
  {Fogedby}},\ }\href {\doibase 10.1103/PhysRevLett.94.195702} {\bibfield
  {journal} {\bibinfo  {journal} {Phys. Rev. Lett.}\ }\textbf {\bibinfo
  {volume} {94}},\ \bibinfo {pages} {195702} (\bibinfo {year}
  {2005})}\BibitemShut {NoStop}%
\bibitem [{\citenamefont {Kelling}\ and\ \citenamefont
  {{\'{O}}dor}(2011)}]{Kelling}%
  \BibitemOpen
  \bibfield  {author} {\bibinfo {author} {\bibfnamefont {J.}~\bibnamefont
  {Kelling}}\ and\ \bibinfo {author} {\bibfnamefont {G.}~\bibnamefont
  {{\'{O}}dor}},\ }\href {\doibase 10.1103/PhysRevE.84.061150} {\bibfield
  {journal} {\bibinfo  {journal} {Phys. Rev. E}\ }\textbf {\bibinfo {volume}
  {84}},\ \bibinfo {pages} {61150} (\bibinfo {year} {2011})}\BibitemShut
  {NoStop}%
\bibitem [{\citenamefont {Pagnani}\ and\ \citenamefont
  {Parisi}(2015)}]{Pagnani}%
  \BibitemOpen
  \bibfield  {author} {\bibinfo {author} {\bibfnamefont {A.}~\bibnamefont
  {Pagnani}}\ and\ \bibinfo {author} {\bibfnamefont {G.}~\bibnamefont
  {Parisi}},\ }\href {\doibase 10.1103/PhysRevE.92.010101} {\bibfield
  {journal} {\bibinfo  {journal} {Phys. Rev. E}\ }\textbf {\bibinfo {volume}
  {92}},\ \bibinfo {pages} {010101(R)} (\bibinfo {year} {2015})}\BibitemShut
  {NoStop}%
\bibitem [{\citenamefont {Carrasco}\ and\ \citenamefont
  {Oliveira}(2022)}]{Ismael22}%
  \BibitemOpen
  \bibfield  {author} {\bibinfo {author} {\bibfnamefont {I.~S.~S.}\
  \bibnamefont {Carrasco}}\ and\ \bibinfo {author} {\bibfnamefont {T.~J.}\
  \bibnamefont {Oliveira}},\ }\href {\doibase 10.1103/PhysRevE.105.054804}
  {\bibfield  {journal} {\bibinfo  {journal} {Phys. Rev. E}\ }\textbf {\bibinfo
  {volume} {105}},\ \bibinfo {pages} {054804} (\bibinfo {year}
  {2022})}\BibitemShut {NoStop}%
\bibitem [{\citenamefont {Kelling}\ \emph {et~al.}(2018)\citenamefont
  {Kelling}, \citenamefont {{\'{O}}dor},\ and\ \citenamefont
  {Gemming}}]{Kelling18}%
  \BibitemOpen
  \bibfield  {author} {\bibinfo {author} {\bibfnamefont {J.}~\bibnamefont
  {Kelling}}, \bibinfo {author} {\bibfnamefont {G.}~\bibnamefont {{\'{O}}dor}},
  \ and\ \bibinfo {author} {\bibfnamefont {S.}~\bibnamefont {Gemming}},\ }\href
  {\doibase 10.1088/1751-8121/aa97f3} {\bibfield  {journal} {\bibinfo
  {journal} {J. Phys. A}\ }\textbf {\bibinfo {volume} {51}},\ \bibinfo {pages}
  {035003} (\bibinfo {year} {2018})}\BibitemShut {NoStop}%
\bibitem [{\citenamefont {Edwards}\ and\ \citenamefont {Wilkinson}(1982)}]{EW}%
  \BibitemOpen
  \bibfield  {author} {\bibinfo {author} {\bibfnamefont {S.~F.}\ \bibnamefont
  {Edwards}}\ and\ \bibinfo {author} {\bibfnamefont {D.~R.}\ \bibnamefont
  {Wilkinson}},\ }\href@noop {} {\bibfield  {journal} {\bibinfo  {journal}
  {Proc. R. Soc. London, Ser. A}\ }\textbf {\bibinfo {volume} {381}},\ \bibinfo
  {pages} {17} (\bibinfo {year} {1982})}\BibitemShut {NoStop}%
\bibitem [{\citenamefont {Villain}(1991)}]{Villain}%
  \BibitemOpen
  \bibfield  {author} {\bibinfo {author} {\bibfnamefont {J.}~\bibnamefont
  {Villain}},\ }\href {\doibase 10.1051/jp1:1991114} {\bibfield  {journal}
  {\bibinfo  {journal} {J. Phys. I France}\ }\textbf {\bibinfo {volume} {1}},\
  \bibinfo {pages} {19} (\bibinfo {year} {1991})}\BibitemShut {NoStop}%
\bibitem [{\citenamefont {Lai}\ and\ \citenamefont {{Das Sarma}}(1991)}]{LDS}%
  \BibitemOpen
  \bibfield  {author} {\bibinfo {author} {\bibfnamefont {Z.-W.}\ \bibnamefont
  {Lai}}\ and\ \bibinfo {author} {\bibfnamefont {S.}~\bibnamefont {{Das
  Sarma}}},\ }\href {\doibase 10.1103/PhysRevLett.66.2348} {\bibfield
  {journal} {\bibinfo  {journal} {Phys. Rev. Lett.}\ }\textbf {\bibinfo
  {volume} {66}},\ \bibinfo {pages} {2348} (\bibinfo {year}
  {1991})}\BibitemShut {NoStop}%
\bibitem [{\citenamefont {Janssen}(1997)}]{Janssen}%
  \BibitemOpen
  \bibfield  {author} {\bibinfo {author} {\bibfnamefont {H.~K.}\ \bibnamefont
  {Janssen}},\ }\href {\doibase 10.1103/PhysRevLett.78.1082} {\bibfield
  {journal} {\bibinfo  {journal} {Phys. Rev. Lett.}\ }\textbf {\bibinfo
  {volume} {78}},\ \bibinfo {pages} {1082} (\bibinfo {year}
  {1997})}\BibitemShut {NoStop}%
\bibitem [{\citenamefont {Halpin-Healy}(2012)}]{healy12}%
  \BibitemOpen
  \bibfield  {author} {\bibinfo {author} {\bibfnamefont {T.}~\bibnamefont
  {Halpin-Healy}},\ }\href {\doibase 10.1103/PhysRevLett.109.170602} {\bibfield
   {journal} {\bibinfo  {journal} {Phys. Rev. Lett.}\ }\textbf {\bibinfo
  {volume} {109}},\ \bibinfo {pages} {170602} (\bibinfo {year}
  {2012})}\BibitemShut {NoStop}%
\bibitem [{\citenamefont {Kim}\ and\ \citenamefont {Kosterlitz}(1989)}]{KK}%
  \BibitemOpen
  \bibfield  {author} {\bibinfo {author} {\bibfnamefont {J.~M.}\ \bibnamefont
  {Kim}}\ and\ \bibinfo {author} {\bibfnamefont {J.~M.}\ \bibnamefont
  {Kosterlitz}},\ }\href@noop {} {\bibfield  {journal} {\bibinfo  {journal}
  {Phys. Rev. Lett.}\ }\textbf {\bibinfo {volume} {62}},\ \bibinfo {pages}
  {2289} (\bibinfo {year} {1989})}\BibitemShut {NoStop}%
\bibitem [{\citenamefont {Forrest}\ and\ \citenamefont
  {Tang}(1990)}]{Forrest90}%
  \BibitemOpen
  \bibfield  {author} {\bibinfo {author} {\bibfnamefont {B.~M.}\ \bibnamefont
  {Forrest}}\ and\ \bibinfo {author} {\bibfnamefont {L.-H.}\ \bibnamefont
  {Tang}},\ }\href {\doibase 10.1103/PhysRevLett.64.1405} {\bibfield  {journal}
  {\bibinfo  {journal} {Phys. Rev. Lett.}\ }\textbf {\bibinfo {volume} {64}},\
  \bibinfo {pages} {1405} (\bibinfo {year} {1990})}\BibitemShut {NoStop}%
\bibitem [{\citenamefont {Kim}(2021)}]{Kim2021}%
  \BibitemOpen
  \bibfield  {author} {\bibinfo {author} {\bibfnamefont {J.~M.}\ \bibnamefont
  {Kim}},\ }\href {\doibase 10.1088/1742-5468/ac0f6f} {\bibfield  {journal}
  {\bibinfo  {journal} {J. Stat. Mech.}\ }\textbf {\bibinfo {volume} {2021}},\
  \bibinfo {pages} {083202} (\bibinfo {year} {2021})}\BibitemShut {NoStop}%
\bibitem [{\citenamefont {Kim}\ and\ \citenamefont {Kim}(2014)}]{Kim2014}%
  \BibitemOpen
  \bibfield  {author} {\bibinfo {author} {\bibfnamefont {S.-W.}\ \bibnamefont
  {Kim}}\ and\ \bibinfo {author} {\bibfnamefont {J.~M.}\ \bibnamefont {Kim}},\
  }\href {\doibase 10.1088/1742-5468/2014/07/P07005} {\bibfield  {journal}
  {\bibinfo  {journal} {J. Stat. Mech.}\ }\textbf {\bibinfo {volume} {2014}},\
  \bibinfo {pages} {P07005} (\bibinfo {year} {2014})}\BibitemShut {NoStop}%
\bibitem [{\citenamefont {Marinari}\ \emph {et~al.}(2000)\citenamefont
  {Marinari}, \citenamefont {Pagnani},\ and\ \citenamefont
  {Parisi}}]{Marinari}%
  \BibitemOpen
  \bibfield  {author} {\bibinfo {author} {\bibfnamefont {E.}~\bibnamefont
  {Marinari}}, \bibinfo {author} {\bibfnamefont {A.}~\bibnamefont {Pagnani}}, \
  and\ \bibinfo {author} {\bibfnamefont {G.}~\bibnamefont {Parisi}},\
  }\href@noop {} {\bibfield  {journal} {\bibinfo  {journal} {J. Phys. A}\
  }\textbf {\bibinfo {volume} {33}},\ \bibinfo {pages} {8181} (\bibinfo {year}
  {2000})}\BibitemShut {NoStop}%
\bibitem [{\citenamefont {Oliveira}(2021)}]{tiago21}%
  \BibitemOpen
  \bibfield  {author} {\bibinfo {author} {\bibfnamefont {T.~J.}\ \bibnamefont
  {Oliveira}},\ }\href {\doibase 10.1209/0295-5075/133/28001} {\bibfield
  {journal} {\bibinfo  {journal} {Europhys. Lett.}\ }\textbf {\bibinfo {volume}
  {133}},\ \bibinfo {pages} {28001} (\bibinfo {year} {2021})}\BibitemShut
  {NoStop}%
\bibitem [{Note1()}]{Note1}%
  \BibitemOpen
  \bibinfo {note} {The ``mixture'' of two lateral sizes ($L$ and $L-1$) is
  needed in some cases (e.g., $d=14$) because $L^d$ is too large, while
  $(L-1)^d$ is relatively small.}\BibitemShut {Stop}%
\bibitem [{\citenamefont {{Gomes-Filho}}\ \emph {et~al.}(2021)\citenamefont
  {{Gomes-Filho}}, \citenamefont {Penna},\ and\ \citenamefont
  {Oliveira}}]{FernandoExp}%
  \BibitemOpen
  \bibfield  {author} {\bibinfo {author} {\bibfnamefont {M.~S.}\ \bibnamefont
  {{Gomes-Filho}}}, \bibinfo {author} {\bibfnamefont {A.~L.~A.}\ \bibnamefont
  {Penna}}, \ and\ \bibinfo {author} {\bibfnamefont {F.~A.}\ \bibnamefont
  {Oliveira}},\ }\href {\doibase 10.1016/j.rinp.2021.104435} {\bibfield
  {journal} {\bibinfo  {journal} {Results in Physics}\ }\textbf {\bibinfo
  {volume} {26}},\ \bibinfo {pages} {104435} (\bibinfo {year}
  {2021})}\BibitemShut {NoStop}%
\bibitem [{\citenamefont {Monthus}\ and\ \citenamefont
  {Garel}(2006)}]{Monthus06}%
  \BibitemOpen
  \bibfield  {author} {\bibinfo {author} {\bibfnamefont {C.}~\bibnamefont
  {Monthus}}\ and\ \bibinfo {author} {\bibfnamefont {T.}~\bibnamefont
  {Garel}},\ }\href {\doibase 10.1103/PhysRevE.74.051109} {\bibfield  {journal}
  {\bibinfo  {journal} {Phys. Rev. E}\ }\textbf {\bibinfo {volume} {74}},\
  \bibinfo {pages} {051109} (\bibinfo {year} {2006})}\BibitemShut {NoStop}%
\bibitem [{\citenamefont {Monthus}\ and\ \citenamefont
  {Garel}(2008)}]{Monthus08}%
  \BibitemOpen
  \bibfield  {author} {\bibinfo {author} {\bibfnamefont {C.}~\bibnamefont
  {Monthus}}\ and\ \bibinfo {author} {\bibfnamefont {T.}~\bibnamefont
  {Garel}},\ }\href {\doibase 10.1088/1742-5468/2008/01/P01008} {\bibfield
  {journal} {\bibinfo  {journal} {J. Stat. Mech.}\ }\textbf {\bibinfo {volume}
  {2008}},\ \bibinfo {pages} {P01008} (\bibinfo {year} {2008})}\BibitemShut
  {NoStop}%
\bibitem [{\citenamefont {Oliveira}\ \emph {et~al.}(2013)\citenamefont
  {Oliveira}, \citenamefont {Alves},\ and\ \citenamefont {Ferreira}}]{tiago13}%
  \BibitemOpen
  \bibfield  {author} {\bibinfo {author} {\bibfnamefont {T.~J.}\ \bibnamefont
  {Oliveira}}, \bibinfo {author} {\bibfnamefont {S.~G.}\ \bibnamefont {Alves}},
  \ and\ \bibinfo {author} {\bibfnamefont {S.~C.}\ \bibnamefont {Ferreira}},\
  }\href@noop {} {\bibfield  {journal} {\bibinfo  {journal} {Phys. Rev. E}\
  }\textbf {\bibinfo {volume} {87}},\ \bibinfo {pages} {040102(R)} (\bibinfo
  {year} {2013})}\BibitemShut {NoStop}%
\bibitem [{\citenamefont {Halpin-Healy}(2013)}]{healy13}%
  \BibitemOpen
  \bibfield  {author} {\bibinfo {author} {\bibfnamefont {T.}~\bibnamefont
  {Halpin-Healy}},\ }\href@noop {} {\bibfield  {journal} {\bibinfo  {journal}
  {Phys. Rev. E}\ }\textbf {\bibinfo {volume} {88}},\ \bibinfo {pages} {042118}
  (\bibinfo {year} {2013})}\BibitemShut {NoStop}%
\bibitem [{\citenamefont {Carrasco}\ \emph {et~al.}(2014)\citenamefont
  {Carrasco}, \citenamefont {Takeuchi}, \citenamefont {Ferreira},\ and\
  \citenamefont {Oliveira}}]{Ismael14}%
  \BibitemOpen
  \bibfield  {author} {\bibinfo {author} {\bibfnamefont {I.~S.~S.}\
  \bibnamefont {Carrasco}}, \bibinfo {author} {\bibfnamefont {K.~A.}\
  \bibnamefont {Takeuchi}}, \bibinfo {author} {\bibfnamefont {S.~C.}\
  \bibnamefont {Ferreira}}, \ and\ \bibinfo {author} {\bibfnamefont {T.~J.}\
  \bibnamefont {Oliveira}},\ }\href@noop {} {\bibfield  {journal} {\bibinfo
  {journal} {New J. Phys.}\ }\textbf {\bibinfo {volume} {14}},\ \bibinfo
  {pages} {123057} (\bibinfo {year} {2014})}\BibitemShut {NoStop}%
\bibitem [{\citenamefont {Alves}\ \emph
  {et~al.}(2014{\natexlab{b}})\citenamefont {Alves}, \citenamefont {Oliveira},\
  and\ \citenamefont {Ferreira}}]{Alves14BD}%
  \BibitemOpen
  \bibfield  {author} {\bibinfo {author} {\bibfnamefont {S.~G.}\ \bibnamefont
  {Alves}}, \bibinfo {author} {\bibfnamefont {T.~J.}\ \bibnamefont {Oliveira}},
  \ and\ \bibinfo {author} {\bibfnamefont {S.~C.}\ \bibnamefont {Ferreira}},\
  }\href {\doibase 10.1103/PhysRevE.90.052405} {\bibfield  {journal} {\bibinfo
  {journal} {Phys. Rev. E}\ }\textbf {\bibinfo {volume} {90}},\ \bibinfo
  {pages} {52405} (\bibinfo {year} {2014}{\natexlab{b}})}\BibitemShut {NoStop}%
\bibitem [{\citenamefont {Almeida}\ \emph {et~al.}(2014)\citenamefont
  {Almeida}, \citenamefont {Ferreira}, \citenamefont {Oliveira},\ and\
  \citenamefont {{Aar{\~{a}}o Reis}}}]{Almeida14}%
  \BibitemOpen
  \bibfield  {author} {\bibinfo {author} {\bibfnamefont {R.~A.~L.}\
  \bibnamefont {Almeida}}, \bibinfo {author} {\bibfnamefont {S.~O.}\
  \bibnamefont {Ferreira}}, \bibinfo {author} {\bibfnamefont {T.~J.}\
  \bibnamefont {Oliveira}}, \ and\ \bibinfo {author} {\bibfnamefont {F.~D.~A.}\
  \bibnamefont {{Aar{\~{a}}o Reis}}},\ }\href@noop {} {\bibfield  {journal}
  {\bibinfo  {journal} {Phys. Rev. B}\ }\textbf {\bibinfo {volume} {89}},\
  \bibinfo {pages} {045309} (\bibinfo {year} {2014})}\BibitemShut {NoStop}%
\bibitem [{\citenamefont {Halpin-Healy}\ and\ \citenamefont
  {Palasantzas}(2014)}]{healy14exp}%
  \BibitemOpen
  \bibfield  {author} {\bibinfo {author} {\bibfnamefont {T.}~\bibnamefont
  {Halpin-Healy}}\ and\ \bibinfo {author} {\bibfnamefont {G.}~\bibnamefont
  {Palasantzas}},\ }\href {\doibase 10.1209/0295-5075/105/50001} {\bibfield
  {journal} {\bibinfo  {journal} {Europhys. Lett.}\ }\textbf {\bibinfo {volume}
  {105}},\ \bibinfo {pages} {50001} (\bibinfo {year} {2014})}\BibitemShut
  {NoStop}%
\bibitem [{\citenamefont {Almeida}\ \emph {et~al.}(2015)\citenamefont
  {Almeida}, \citenamefont {Ferreira}, \citenamefont {Ribeiro},\ and\
  \citenamefont {Oliveira}}]{Almeida15}%
  \BibitemOpen
  \bibfield  {author} {\bibinfo {author} {\bibfnamefont {R.~A.~L.}\
  \bibnamefont {Almeida}}, \bibinfo {author} {\bibfnamefont {S.~O.}\
  \bibnamefont {Ferreira}}, \bibinfo {author} {\bibfnamefont {I.~R.~B.}\
  \bibnamefont {Ribeiro}}, \ and\ \bibinfo {author} {\bibfnamefont {T.~J.}\
  \bibnamefont {Oliveira}},\ }\href@noop {} {\bibfield  {journal} {\bibinfo
  {journal} {Eur. Lett.}\ }\textbf {\bibinfo {volume} {109}},\ \bibinfo {pages}
  {46003} (\bibinfo {year} {2015})}\BibitemShut {NoStop}%
\bibitem [{\citenamefont {Almeida}\ \emph {et~al.}(2017)\citenamefont
  {Almeida}, \citenamefont {Ferreira}, \citenamefont {Ferraz},\ and\
  \citenamefont {Oliveira}}]{Almeida17}%
  \BibitemOpen
  \bibfield  {author} {\bibinfo {author} {\bibfnamefont {R.~A.~L.}\
  \bibnamefont {Almeida}}, \bibinfo {author} {\bibfnamefont {S.~O.}\
  \bibnamefont {Ferreira}}, \bibinfo {author} {\bibfnamefont {I.}~\bibnamefont
  {Ferraz}}, \ and\ \bibinfo {author} {\bibfnamefont {T.~J.}\ \bibnamefont
  {Oliveira}},\ }\href@noop {} {\bibfield  {journal} {\bibinfo  {journal} {Sci.
  Rep.}\ }\textbf {\bibinfo {volume} {7}},\ \bibinfo {pages} {3773} (\bibinfo
  {year} {2017})}\BibitemShut {NoStop}%
\end{thebibliography}%

\end{document}